\documentclass[usenatbib,referee]{mn2e}
\usepackage[utf8]{inputenc}
\usepackage{graphicx}
\usepackage{epsfig}
\usepackage{hyperref}
\usepackage{amsmath}
\usepackage{amsfonts}
\usepackage{amssymb}

\title{Spin Angular Momentum  Evolution of the Long Period Algols}
\author[A. Dervi\c{s}o\u{g}lu et al.]{A. Dervi\c{s}o\u{g}lu$^{1,2}$\thanks{E-mail: 
ahmetdervisoglu@mail.ege.edu.tr}, Christopher
A. Tout$^{2}$ and C. \.{I}bano\u{g}lu$^{1}$\\ 
$^{1}$Ege University, Science Faculty, Astronomy and Space Sciences Dept., 35100 Bornova, \.{I}zmir, Turkey\\
$^{2}$Institute of Astronomy, The
Observatories, Madingley Road, Cambridge CB3 0HA} 
\begin{document}

\date{Accepted 200x December 15. Received 200x December 14; in
original form 1988 October 11} 
\pagerange{\pageref{firstpage}--\pageref{lastpage}} \pubyear{2009}

\maketitle
\label{firstpage}

\begin{abstract}
We consider the spin angular momentum evolution of the accreting
components of Algol-type binary stars.  In wider Algols the accretion
is through a disc so that the accreted material can transfer enough
angular momentum to the gainer that material at its equator should be
spinning at break-up.  We demonstrate that even a small amount of mass
transfer, much less than required to produce today's mass ratios,
transfers enough angular momentum to spin the gainer up to this
critical rotation velocity.  However the accretors in these systems
have spins typically between $10$ and~$40\,$per cent of the critical
rate.  So some mechanism for angular momentum loss from the gainers is
required.  Unlike solar type chromospherically active stars, with
enhanced magnetic activity which leads to angular momentum and mass
loss, the gainers in classical Algols have radiative envelopes.  We
further find that normal radiative tides are far too weak to account
for the necessary angular momentum loss.  Thus enhanced mass loss in a
stellar wind seems to be required to spin down the gainers in
classical Algol systems.  We consider generation of magnetic fields in
the radiative atmospheres in a differentially rotating star and the
possibility of angular momentum loss driven by strong stellar
winds in the intermediate mass stars, such as the primaries of the
Algols.  Differential rotation, induced by the accretion itself, may
produce such winds which carry away enough angular momentum to reduce
their rotational velocities to the today's observed values.  We apply
this model to two systems with initial periods of 5\,d, one with
initial masses $5$ and $3\,\rm{M}_{\odot}$ and the other with $3.2$
and $2\,\rm{M}_{\odot}$.  Our calculations show that, if the mass
outflow rate in the stellar wind is about $10\,$per cent of the
accretion rate and the dipole magnetic field is stronger than about
$1\,$kG, the spin rate of the gainer is reduced to below break-up
velocity even in the fast phase of mass transfer.  Larger mass loss is
needed for smaller magnetic fields.  The slow rotation of the gainers
in the classical Algol systems is explained by a balance between the
spin-up by mass accretion and spin-down by a stellar wind linked to a
magnetic field.

\end{abstract}

\section{Introduction}

Evolution of single stars is now well modelled \citep[see for
  example][]{pols1995}.  There remain concerns with mass loss,
rotation and convection but appropriate and successful empirical
treatments exist.  Evolution of a binary star has several additional
complications associated with interaction between the components.
Since solving the mystery of Algol systems
\citep{hoyle1955,crawford1955}, the prototype of semi-detached
Algol-type binary stars with one evolved and one main-sequence
component, we realize that there are some stages of evolution when
interaction between the components is unavoidable.  We must therefore
take into account, in our calculations, the mass transfer and mass
loss together with any angular momentum and magnetic interaction
between the components, at least in some critical phases, to fully
understand evolution of a binary system.

Over the last few decades, the evolution of Algols has been modelled
with well defined approximations such as conservation of total mass
and total orbital angular momentum.  The effect of mass transfer on
the structure of both stars can be modelled reasonably well.  The
angular momentum transfer during mass exchange, however, is not well
understood.  As we shall see in section~\ref{accdiscs} there are some
episodes of mass transfer in Algols when accretion discs or disc-like
structures form around the mass gainer.

Current approximations of binary star evolution do not adequately
explain the spin angular momentum of the mass-gaining components
because the high specific angular momentum of the disc material should
easily spin these stars up to their critical break-up rotational
velocities in less than the time needed to reverse the mass ratio of
system and enter the Algol phase.  Here we discuss formation of discs
in classical Algol systems and consider the spin angular momentum
evolution of mass accreting components, taking into account discs,
tides and magnetic stellar winds.  We demonstrate that tidal effects
play a minor role in the removal of excess angular momentum
from the gainers and rely on a magnetically locked stellar wind to do
this.

\begin{table*}

\caption{The absolute parameters of Algol primary components
  \citep{ibanoglu2006}.  For each star, columns $2-8$ are the orbital
  period, mass ratio, masses, radii and inclination.  Then $v_{\rm
    syn}$ would be the equatorial velocity of the mass gainer (star~1)
  if it were synchronous while $v_{\rm eq}\sin i$ is the measured
  projected velocity and $F = v_{\rm eq}/v_{\rm syn}$.
  \label{table}}
\begin{tabular}{lccccccccccr}

\hline
Name	&$P/\rm d$	&$q$	&$M_{1}/\rm{M}_{\odot}$	&$M_{2}/\rm{M}_{\odot}$	&$R_{1}/\rm{R}_{\odot}$	&$R_{2}/\rm{R}_{\odot}$	&$i$ 	&$v_{\rm syn} \sin i$ 	&$v_{\rm eq} \sin i$	&$F$	&Ref.\\ 
\hline
TW And	&4.12	&0.210	&1.68	&0.32	&2.19	&3.37	&87	&27	&32	&1.19	&5\\
KO Aql	&2.86	&0.217	&2.53	&0.55	&1.74	&3.34	&78	&30	&41	&1.37	&4\\
IM Aur	&1.25	&0.311	&2.24	&0.76	&2.57	&1.74	&75	&101	&135	&1.34	&1\\
R CMa	&1.14	&0.170	&1.07	&0.17	&1.50	&1.15	&80	&76	&98	&1.29	&1\\
S Cnc	&9.48	&0.090	&2.51	&0.23	&2.15	&5.25	&83	&12	&174	&14.5	&5\\
RZ Cas	&1.20	&0.351	&2.10	&0.74	&1.67	&1.94	&83	&62	&87	&1.40	&3\\
TV Cas	&1.81	&0.470	&3.78	&1.53	&3.15	&3.29	&79	&90	&79	&0.88	&3\\
U Cep	&2.49	&0.550	&3.57	&1.86	&2.41	&4.40	&88	&56	&437	&7.80	&3\\
RS Cep	&12.4	&0.145	&2.83	&0.41	&2.65	&7.63	&87	&11	&170	&15.4	&1\\
XX Cep	&2.34	&0.150	&2.03	&0.33	&2.12	&2.25	&85	&46	&47	&1.02	&2\\
U CrB	&3.45	&0.289	&4.74	&1.46	&2.79	&4.83	&82	&48	&60	&1.25	&1\\
SW Cyg	&4.57	&0.190	&2.50	&0.50	&2.60	&4.30	&83	&22	&196	&8.91	&3\\
WW Cyg	&3.32	&0.310	&2.10	&0.60	&2.00	&7.00	&89	&31	&41	&1.32	&2\\
TW Dra	&2.81	&0.470	&1.70	&0.80	&2.40	&3.40	&86	&43	&37	&0.86	&2\\
AI Dra	&1.20	&0.429	&2.86	&1.34	&2.17	&2.42	&78	&83	&85	&1.02	&1\\
S Equ	&3.44	&0.130	&3.24	&0.42	&2.74	&3.24	&87	&40	&52	&1.30	&4\\
AS Eri	&2.66	&0.110	&1.92	&0.21	&1.57	&2.19	&80	&29	&36	&1.02	&5\\
RX Gem	&12.2	&0.254	&4.40	&0.80	&4.80	&7.00	&85	&20	&157	&7.85	&2\\
RY Gem	&9.30	&0.193	&2.04	&0.39	&2.38	&6.19	&83	&13	&70	&5.38	&5\\
AD Her	&9.77	&0.350	&2.90	&0.90	&2.60	&7.70	&84	&13	&143	&11.0	&2\\
TT Hya	&6.95	&0.224	&2.63	&0.59	&1.95	&5.87	&84	&15	&164	&10.9	&2\\
$\delta$ Lib	&2.33	&0.345	&4.70	&1.70	&4.12	&3.88	&81	&89	&68	&0.76	&1\\
AU Mon	&11.1	&0.199	&5.93	&1.18	&5.28	&10.04	&79	& 24	&124	&5.17	&5\\
TU Mon	&5.09	&0.210	&12.60	&2.70	&5.60	&7.10	&89	&56	&153	&2.73	&2\\
AT Peg	&1.15	&0.484	&2.50	&1.21	&1.91	&2.11	&76	&80	&82	&1.02	&1\\
$\beta$ Per	&2.87	&0.217	&3.70	&0.81	&2.74	&3.60	&82	&51	&52	&1.02	&3\\
RW Per	&14.2	&0.150	&2.56	&0.38	&2.80	&7.30	&81	&10	&161	&16.1	&3\\
RY Per	&6.86	&0.271	&6.24	&1.69	&4.06	&8.10	&83	&30	&213	&7.10	&1\\
Y Psc	&3.77	&0.250	&2.80	&0.70	&3.06	&3.98	&87	&37	&38	&1.03	&3\\
RZ Sct	&15.2	&0.216	&5.50	&1.50	&11.00	&14.00	&83	&36	&222	&6.17	&3\\
V356 Sgr	&8.89	&0.380	&12.20	&4.70	&8.50	&15.40	&85	&48	&212	&4.42	&1\\
V505 Sgr	&1.18	&0.520	&2.68	&1.23	&2.24	&2.17	&80	&85	&101	&1.19	&1\\
U Sge	&3.38	&0.370	&4.45	&1.65	&4.00	&5.00	&90	&60	&76	&1.27	&3\\
$\lambda$ Tau	&3.95	&0.263	&7.18	&1.89	&6.40	&5.30	&76	&80	&88	&1.10	&1\\
TX Uma	&3.06	&0.248	&4.76	&1.18	&2.83	&4.24	&82	&46	&63	&1.37	&2\\
Z Vul	&2.45	&0.430	&5.40	&2.30	&4.30	&4.50	&89	&89	&135	&1.52	&1\\

\hline

\end{tabular}
\begin{list}{}{}
\item[References:]{\small (1) \cite{vanhamme90} , (2) \cite{etzel93},(3) \cite{mukher96}, 
(4) \cite{soydugan2007} and (5) \cite{glazunova2008}.}
\end{list}
\end{table*}

\section{Observations and Motivation}

\begin{figure}
\includegraphics[width=8.5 cm]{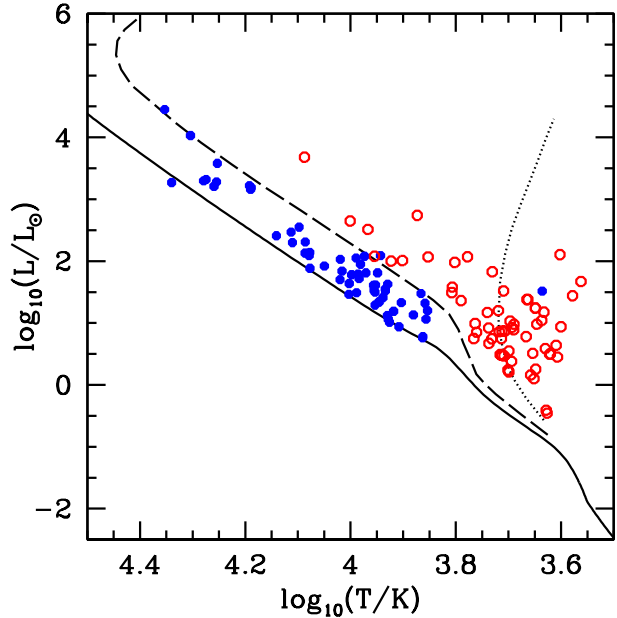}
\caption{Locations of the components of the well known Algol systems
  in a Hertzsprung--Russell diagram. Dots and circles are
  the primary and secondary components, respectively.  The zero-age
  main sequence (ZAMS, continuous), terminal-age main sequence (TAMS, dashed) and base of
  the giant branch (BGB, dotted) are as found by
  \citet{pols1998}. \label{figalgolhr}}
\end{figure}

Recently \citet{ibanoglu2006} compiled
and analysed 61~Algols for which the fundamental parameters are well
known.  In Fig.~\ref{figalgolhr} we show locations of the primary and
secondary components in the Hertzsprung-Russell diagram (HRD).
We use the observers' convention and refer to the brighter, hotter and
currently more massive star as the primary component with mass $M_1$
and the redder, mass losing component as the secondary with mass $M_2$.
\citet{ibanoglu2006} arrived at several interesting observations
concerning the relation between orbital angular momentum and mass.
\begin{enumerate}

\item Semi-detached binaries (SDBs) with mass ratios $q = M_2/M_1 >
  0.3$ and orbital periods of $P>5\,$d have almost the same angular
  momentum as detached binaries (DBs).  However the SDBs with short
  periods have lower angular momentum even when they have the same
  mass ratios.

\item The orbital angular momenta of SDBs with periods $P<5\,$d and
  $P>5\,$d are $45$~and $25\,$per cent smaller, respectively, than
  those of DBs of a total mass of about $3\,{\rm M}_\odot$.

\item The secondaries of SDBs with orbital periods longer than $5\,$d
  have angular momenta twice that of secondaries with the same mass
  but with a period shorter than $5\,$d.

\item The specific angular momenta of systems with $P>5\,$d are about
  $24\,$per cent larger than those of the systems with $P<5\,$d for
  primary components of the same mass.  More extremely the specific
  angular momenta of the longer period systems are $65\,$ per cent
  greater than those of the shorter period systems with the same mass
  secondary star.

\end{enumerate}

\begin{figure}
\includegraphics[width=8.5 cm]{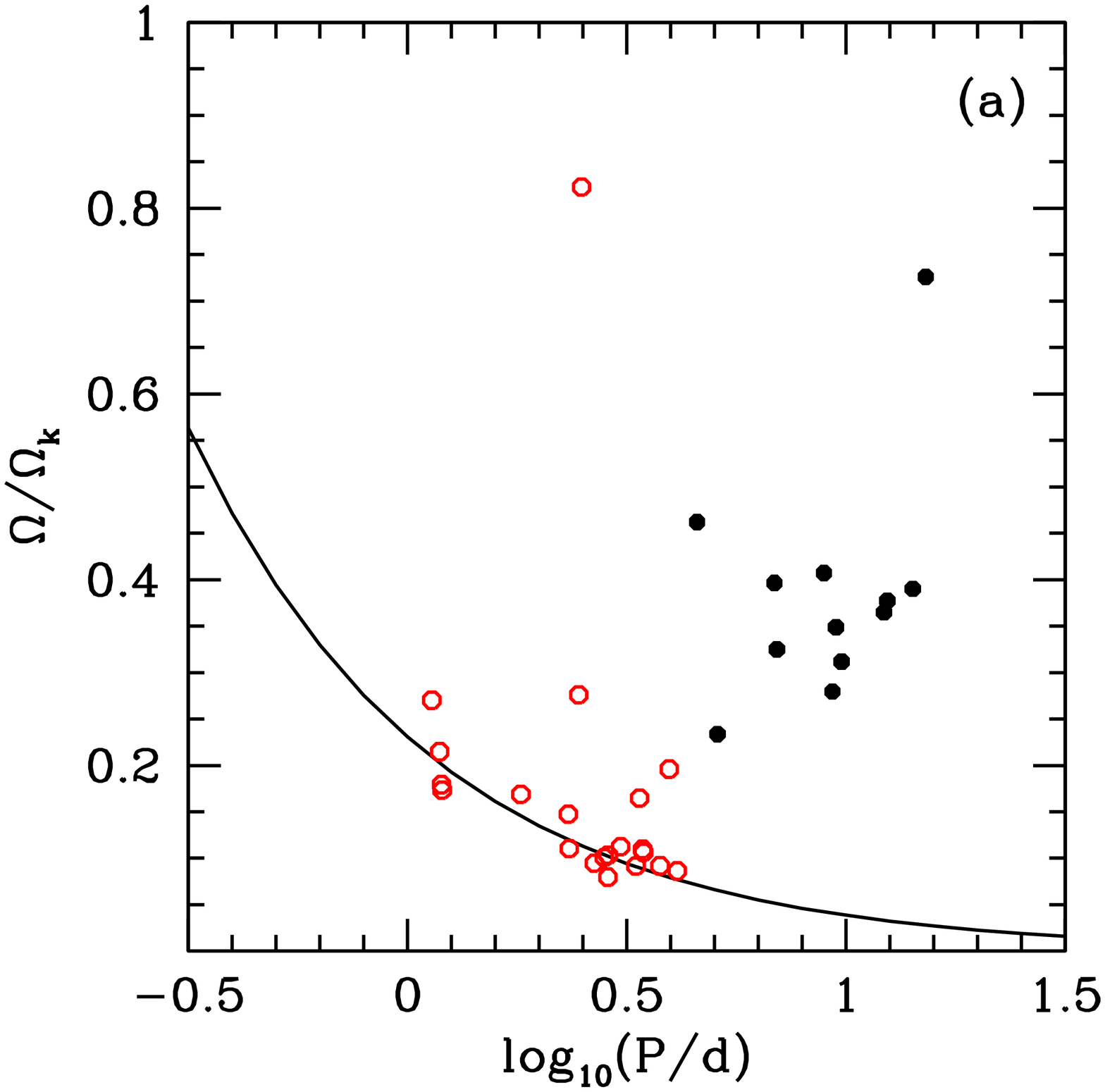}
\includegraphics[width=8.5 cm]{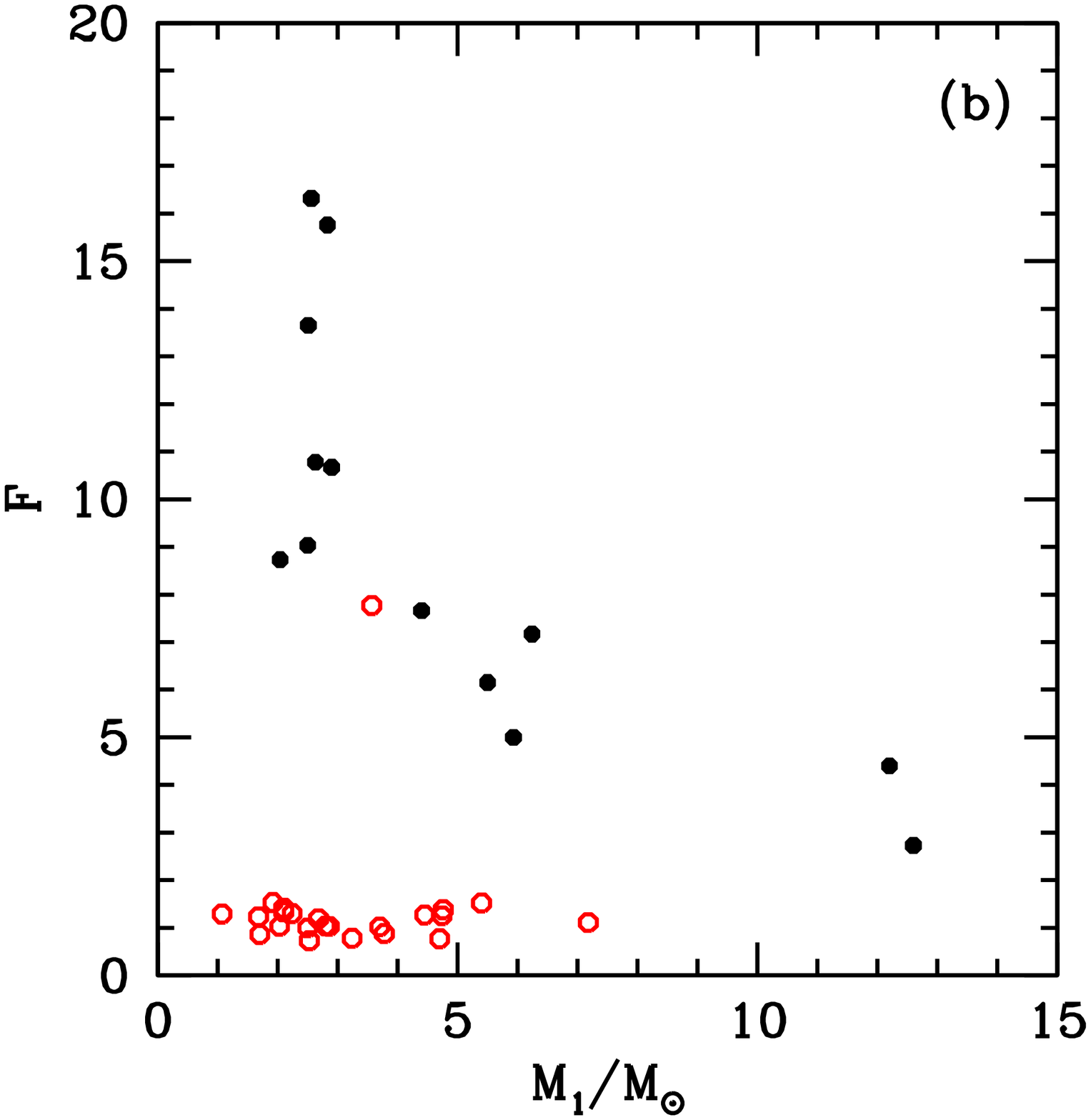}
\caption{(a) The ratios of spin angular to Keplerian angular
  velocities for the mass accreting primaries of the semi-detached
  binaries are plotted against their orbital periods.  The solid line
  is the synchronous value at that period.  (b) The ratios of the
  observed equatorial to the computed synchronous rotational
  velocities, $\rm{F}= v_{\rm eq}/ v_{\rm syn}$, for the same stars
  are plotted against the mass of the gainer.  Open circles and solid
  dots represent the SDBs with periods $P<5\,$d and $P>5\,$d,
  respectively.  There is a marked difference above and below $P =
  5\,$d.  At shorter periods the accretors are close to synchronous,
  while at longer periods they tend to spin somewhat faster.
\label{figprot}}
\end{figure}

These results suggest that different mechanisms govern angular
momentum evolution for short and long period SDBs.  Absolute
parameters and projected rotational velocities measured with great
accuracy are given for some Algols in Table~\ref{table}.  The spins of
the primary stars in SDBs show a marked distinction between short and
long period systems at an orbital period of about $5\,$d
\citep{eggleton2006} as we show in Fig.~\ref{figprot}.  This indicates
that the observed rotational velocities of the primaries are far from
synchronized with the orbit in longer period systems, $P>5\,$d.  The
main discriminator of the high rotational velocity of a gainer in a
classical Algol appears to be its orbital period, independent of its
mass.  However, if we consider only long period Algols, the spin rate
appears to decrease as the mass of the gainer increases.  It should be
noted that the gainer in U~Cep seems to rotate faster than other
short-period Algols despite its $2.5\,$d orbital period.  Photometric
and spectroscopic observations indicate that this system shows
transient disc because the eclipse durations vary from time to time
\citep{gimenez1996,manzoori2008}.  It shows variations in both orbital
period and total luminosity consistent with mass transfer and
convective activity.  This slightly anomalous behaviour is probably
due to the mass ratio in U~Cep being very close to the critical mass
ratio at which mass transfer proceeds dynamically \citep{tout1991}.

The above results led us to reconsider tidal interaction and angular
momentum transfer in systems in which mass transfer is still
occurring.  A number of angular momentum loss mechanisms
\citep{packet1981, eggleton2000, chen2006} have been suggested to
explain the evolution of Algols but none is entirely satisfactory.  In
any case, it is well established that accretion discs can be formed in
phases of evolution when the relative radius of the mass accreting
star is small enough.

\subsection{Accretion Discs}
\label{accdiscs}

\begin{figure}
\includegraphics[width=8.5 cm]{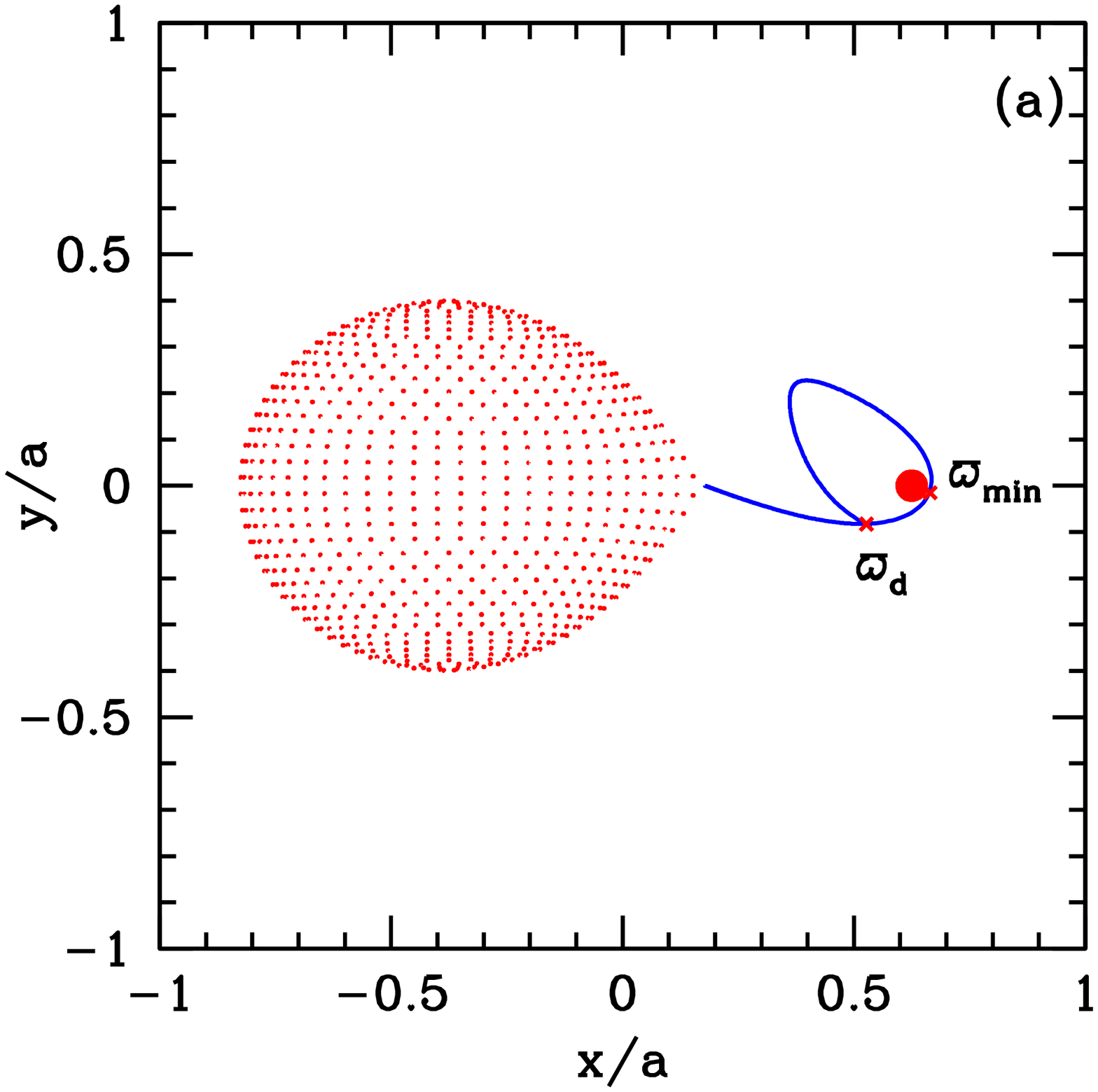}
\includegraphics[width=8.5 cm]{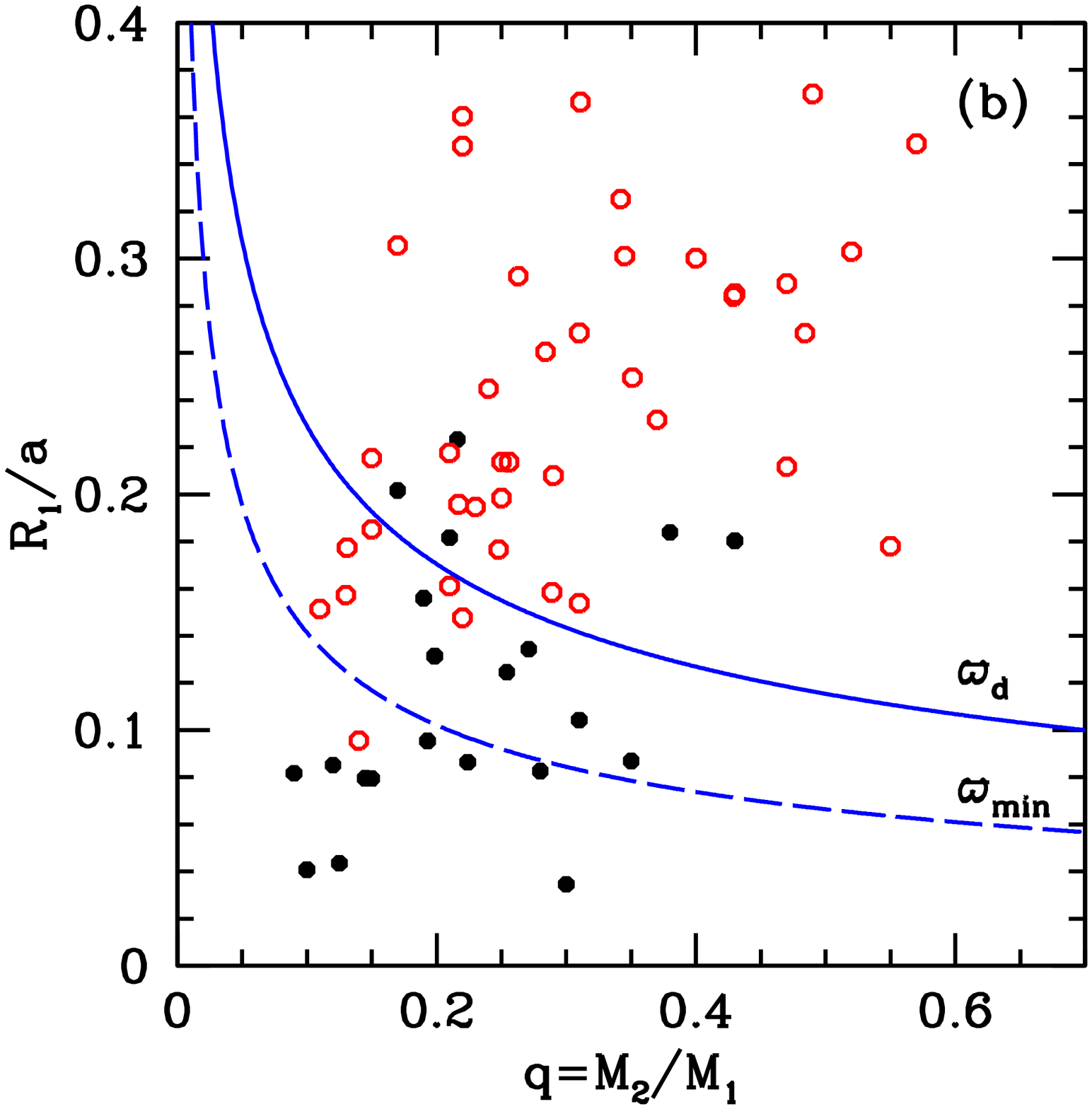}
\caption{
\label{figlubow}
(a) The orbit of the mass flow in a semi-detached system with masses
$5\,\rm{M}_{\odot}$ and $3\,\rm{M}_{\odot}$ and an orbital period of
$5\,$d.  The coordinates $x$ and $y$ are scaled to the orbital
separation $a$ and are in the the equatorial plane of the binary,
perpendicular to the axis of rotation and in a frame corotating with
the system.  The mass stream trajectory is calculated with
\citet{flannery1975}'s approximations.  (b) The positions of all well
known Algol primary stars in a fractional radius against mass-ratio
diagram.  The radii below which a disc must form $\varpi_{\rm min}$
and below which a disc may form $\varpi_{\rm d}$ are indicated.  The
solid dots are the gainers with permanent accretion discs among the
long period Algols.}
\end{figure}

Classical Algols are semi-detached interacting eclipsing binary stars
in which the less massive, evolved secondary component (spectral
type~F or later~G and luminosity class of giant or sub-giant) has
expanded enough to fill its Roche lobe.  These less massive cool
secondaries are transferring material through a gas stream on to a
B~or A~spectral-type main-sequence primary component.  In the long
period, $P>5\,$d, Algols the mass gaining components are small enough,
relative to the binary separation, that mass transfer takes place
through an accretion disc.  The in-falling material has too much
angular momentum for the stream to directly impact on the accretor.
\citet{lubow1975} examined the condition for the formation of discs in
semi-detached systems by modelling the stream as a ballistic flow from
the inner Lagrangian point.  As seen in Fig.~\ref{figlubow}a
\citep[similar to fig.~4 of][]{lubow1975}, if the minimum distance of
the stream from centre of the gainer $a\varpi_{\rm \min}$ is smaller
than the radius of detached component ($R_{1}$) the transferring
material can impact directly on its surface.  This impact leads to the
formation of variable accretion structures.  Otherwise, when
$a\varpi_{\rm \min}>R_{1}$, the mass flow misses the star and collides
with itself at a larger radius.  Radial motion is dissipated and the
resulting ring of material spreads viscously to form a permanent
accretion disc of radius $a\varpi_{\rm \rm d}$.  If the accretor has a
radius between $a\varpi_{\rm min}$ and $a\varpi_{\rm d}$ a disc could
exist because it could intercept the stream before it impacts the
star.  In such a case the disc could be transient.  If $R_1 >
a\varpi_{\rm d}$ then the stream must impact the star directly.
Expressing these distances relative to the separation $a$, we can
predict the presence of discs in semi-detached systems.  Both
$\varpi_{\rm \min}$ and $\varpi_{\rm \rm d}$ are functions only of
mass ratio $q$.  The majority of Algols with $P > (4-5)\,$d are within
the region (Fig.~\ref{figlubow}b) where we expect stars to have either
a permanent or transient disc.

Indeed the presence of a disc around the gainer is easily confirmed with
optical spectra of the primary star.  \citet{richards1999} observed
and analysed optical spectra of Algols and demonstrated that
those with periods $P>(4-5)\,$d show double-peaked H$\alpha$
emission, characteristic of accretion discs, while those of shorter
periods show only single-peaked emission, characteristic of a mass
stream structure \citep[fig.~3 of][]{richards1999}.

It is thus well established, both observationally and theoretically,
that long period Algols develop a permanent or transient accretion
disc or disc-like structure during mass transfer.  In the next
section we examine angular momentum transport and loss mechanisms to
explain the asynchronous rotational velocities of detached components
in the presence of discs or disc-like structures.

\section{Models} \label{models}

In accretion disc theory, the in-falling matter first forms
an accretion ring.  \citet{stothers1972} proposed a model
in which the accreting star is driven into differential rotation by
the presence of such a ring or disc around it.  Viscous forces transfer most
of the angular momentum (AM) to the outer edge of the ring while mass falls
inwards.  The ring spreads out.  Eventually a disc forms and this
allows the matter at the inner edge to fall on to the surface of the star.
For such a Keplerian disc, the angular velocity $\Omega_{\rm k}$ of
material at radius $R$ is given by
\begin{equation}
\Omega_{\rm k}^{2}=\frac{GM}{R^{3}},
\label{eqkepler}
\end{equation}
where $G$ is Newton's gravitational constant and the $M$ is the mass
of the accreting star.  Hence, the specific angular momentum of
accreted material at the surface of the star, of radius $R$, is
\begin{equation}
h_{\rm d}=\sqrt{GMR}.
\end{equation}
This specific angular momentum of accreted material would be equal to
that at the equator of the gainer if it were critically rotating at
brake up.  It is much larger than found in normal stars.  When accreting
at a rate $\dot{M}_{\rm acc}$ the rate of angular momentum transferred
from the disc to the star is
\begin{equation} \label{Tacc}
\frac{dJ_{\rm acc}}{dt}=\dot{M}_{\rm acc}\sqrt{GMR}.
\end{equation}

\begin{figure}
\begin{centering}
\includegraphics[width=8.5 cm]{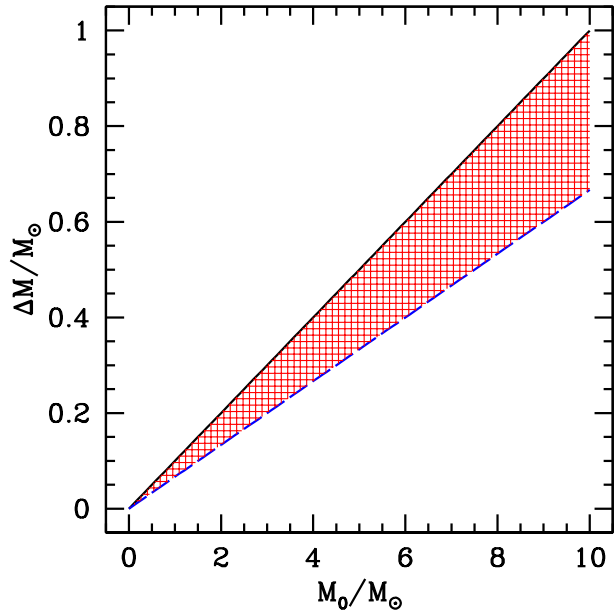} 
\par\end{centering}
\caption{The amount of accreted mass needed to spin a star with
  initial mass of $M_{0}/\rm{M}_{\odot}$ up to its critical rate
  $\Omega_{\rm k}$ when initially $\Omega_0 = 0.1\Omega_{\rm k}$
  (solid line) and $0.4\Omega_{\rm k}$ (dashed line).
\label{figacc}}
\end{figure}

Assuming a negligible change in stellar radius, we can determine the
amount of mass $\Delta M$ that must be transferred through the disc to
spin the star up to its critical angular velocity $\Omega_{\rm k}$
from an initial $\Omega_0$ when it had mass $M_0$.  Let the radius of
gyration of the star be $kR$ so that its total angular momentum is
$k^2MR^2\Omega$ when spinning rigidly at $\Omega$ then
\begin{equation} \label{eqdelm}
\Delta M=\frac{k^{2}}{1-k^{2}}\left(1-\frac{\Omega_0}{\Omega_{\rm k}}\right)M_0.
\end{equation}
A more precise formula was derived by \citet{packet1981} who took
account of the change in the mass of the star but this is unnecessary
for our purposes because $\Delta M$ is always small.  For
main-sequence stars $k^{2}\approx 0.1$ and varies little.  Thus when
$0.1 < \Omega_0/\Omega_{\rm k} < 0.4$ we find $0.1 > \Delta M/M_0 >
0.06$.  This is very small when we consider that all classical Algols
have a mass ratio of $q < 0.7$ \citep[mostly $q \approx 0.2$ according
  to][]{ibanoglu2006} which indicates that the losers in the classical
Algols have transferred more on the order of $1\,\rm{M}_\odot$.  The shaded
area in Fig.~\ref{figacc} shows the amount of material that must be
accreted from a disc to spin the star up to its critical rotational
velocity.  Despite having high spin velocities, $0.1 <
\Omega/\Omega_{\rm k} < 0.4$, observations show that the detached
components in most of the Algols do not actually attain their critical
rotational velocity (Table~\ref{table}, Fig.~\ref{figprot}a).  The
only exception, with the high ratio of $\Omega/\Omega_{\rm k} = 0.72$, is
RZ~Sct.  This star's radial velocity curve is distorted
\citep[e.g][]{mcnamara1957}.  It shows emission in $H_\alpha$ outside
eclipse \citep{mcnamara1957,hansen1959} and its light curve displays
distortions due to an accretion stream \citep{olson1989}.  These
observed phenomena may be taken as the signature of a rapid mass
transfer phase.

In all cases a mechanism is needed to dissipate this excess angular
momentum, along with associated energy.  Here we examine various
mechanisms for angular momentum loss and compare with the
observations.

\subsection{Tidal Forces and Energy Dissipation Mechanisms}

Tidal forces have been quantified since the work of \citet{darwin1879}
and are, perhaps, the most well understood mechanism for angular
momentum transfer within binary stars.  Tidal interactions exchange
angular momentum between the orbit and stellar spins by torques.  They
act to synchronize stellar spins with the orbital period.  According
to \citet{zahn2005} the time-scale for synchronization $t_{\rm sync}$
is given by
\begin{equation}
\frac{1}{t_{\rm sync}} = -\frac{1}{\Omega-\omega}\dot{\Omega}
\approx \frac{1}{t_{\rm diss}}q^{2}\frac{MR^{2}}{I}\left(\frac{R}{\rm
a}\right)^{6},
\label{eqsynch}
\end{equation}
where $\omega$ is orbital angular velocity, $a$ is separation of
system, $I = k^2MR^2$ is the moment of inertia of the primary star and
$t_{\rm diss}$ is the time-scale for the most effective dissipation
mechanism.  It depends on the stellar type.  For stars with convective
envelopes the kinetic energy of the equilibrium tides is dissipated by
turbulent convective eddies.  For those with radiative envelopes the
shortest time-scale is through gravity wave dissipation.  It is well
known that the energy dissipation in convective envelopes is much more
effective than in radiative envelopes \citep{zahn2005}.

\begin{figure}
\begin{centering}
\includegraphics[width=8.5 cm]{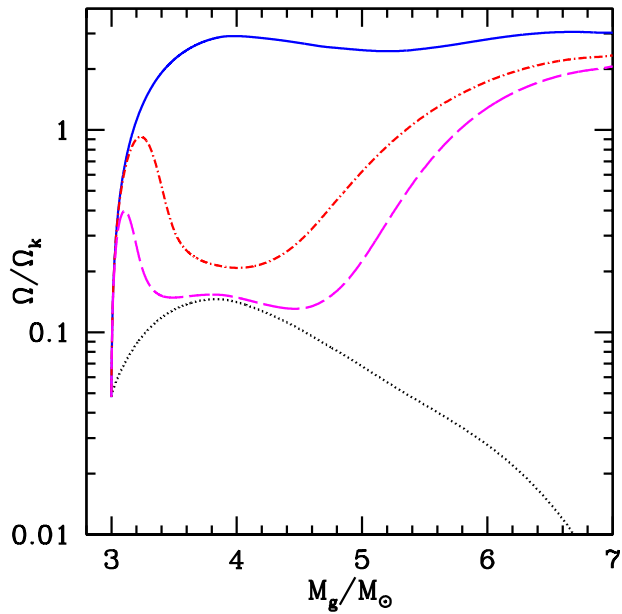}
\par \end{centering}
\caption{Spin evolution of a gainer in a binary with initial masses of
  $5 + 3\,\rm{M}_{\odot}$ and a period of $P=5\,$d in the case when
  total mass and angular momentum are conserved.  Calculations were
  made for standard tides (equation~\ref{eqsynch}, solid line) and
  tides increased by factors of $10^7$ (dot-dashed line) and $10^8$
  (dashed line).  The fully synchronous case (dotted line) is plotted
  for comparison. \label{figtidal}}
\end{figure}

As we shall describe in the details of the binary star evolution
models in section~\ref{magwind}, we have examined the efficiency of
the tidal torque at opposing the spinning up effect of accreted
material from a disc.  Gainers in Algol systems are early type stars,
so we take the approximation of the dynamical tide with radiative
damping given by \citet{hurley2002} to evaluate $t_{\rm diss}$ for
equation~\ref{eqsynch}.  In the classical evolution of an Algol the
initially more massive star evolves more quickly, overfills its Roche
lobe and begins a phase of rapid mass transfer soon after it departs
from the main sequence as it rapidly becomes a giant star.  The
primary star is at this time relatively compact and substantially
smaller than its companion so that an accretion disc is likely to
form.  When the accretion is conservative and uniform over time, so
that the rate of accretion from the disc to the gainer ($\dot{M}_{\rm
  acc}$) is the same as the mass-loss rate from the donor
($\dot{M}_{2}$), we can calculate the spin angular momentum gained
during mass transfer.  Initially the angular momentum of the gainer is
\begin{equation}
J_{\rm s0}=k^2MR^{2}\Omega_{0}\,.
\end{equation}
The net change depends on the competition between the torque of
accreted material (equation~\ref{Tacc}) and the opposing spin-down torque
of tides obtained from equation~\ref{eqsynch}, by
\begin{equation} \label{eqtide}
\left(\frac{dJ}{dt}\right)_{\rm tid}=k^2MR^{2}\dot{\Omega},
\end{equation}
where $k$ remains approximately constant.  In Fig.\,\ref{figtidal} we
show the variation in the angular velocity of a gainer undergoing mass
accretion through a disc with its increasing mass.  We used a detailed
evolution model for a system of stars initially of masses $5$~and
$3\,\rm{M}_{\odot}$ and an orbital period of $P=5\,$d.  We allow the
star to rotate faster than its break-up rate for illustration only.
We find tides are almost incapable of synchronizing the star with the
orbit because the gainer reaches brake-up velocity after accreting
only a small amount of matter as indicated by equation~\ref{eqdelm}.
We found that, to affect the star's spin, the tides would need to be
stronger by more than a factor of $10^7$.  There is no known physical
basis for this.  We might also hope that a convective core's ability
to dissipate energy by tidal forces may have an effect but the strong
dependence of the synchronization time on $(R/a)^6$ in
equation~\ref{eqsynch} means that the contribution of the convective
core is too small and may be neglected.  Thus we conclude that tides
are insufficient to synchronize the stellar spin of the gainer with
the orbit when accretion is through a disc.

\subsection{Magnetic Winds \label{magwind}}

Our analysis has shown that tides are insufficient to dissipate the
excess angular momentum accreted by Algol primaries with discs.  So we
consider alternatives that are applicable to this case.  The total
angular momentum lost from a star in a wind coupled to a magnetic
field is equivalent to angular momentum carried away by the wind
material co-rotating up to the Alfv\'{e}n surface
\citep{weber1967,mestel1968}.  So the rate of change of angular
momentum of the star owing to the wind is
\begin{equation}\label{meq1}
\left(\frac{dJ}{dt}\right)_{\rm w} = \dot{M}_{\rm w} R_{\rm A}^2 \Omega ,
\end{equation}
where $\Omega$ is the angular velocity of the star and $R_{\rm
  A}$ is the Alfv\'{e}n radius at which outflow speed equals the local
magnetic Alfv\'{e}n speed
\begin{equation}\label{meq2}
v_{\rm A} \approx \dfrac{B_{\rm A}}{\sqrt{4\pi \rho_{\rm A}}}, 
\end{equation}
where $\rho _{\rm A}$ is the density of wind material, $B_{\rm A}$ is
the magnetic field strength at the Alfv\'{e}n surface and $\dot{M}_{\rm
  w} < 0$ is the mass-loss rate.  For a spherical outflow
\begin{equation}\label{meq3}
\rho_{\rm A} \approx \frac{-\dot{M}_{\rm w}}{4\pi R_{\rm A}^2 v_{\rm A}}  .
\end{equation}

The angular momentum loss rate depends on the field structure and flow
velocity which are not easily determined a priori for the wind.  We
need to make assumptions about both.  To model the magnetic field
structure in a simple manner, we assume that field strength
follows a single power law of the form
\begin{equation}\label{meq4}
B_{\rm A} = B_{\rm s}(R/R_{\rm A})^n, 
\end{equation}
where $n$ describes the geometry of the stellar field and $n=3$
corresponds to a dipole field \citep{weber1967,mestel1987} while $
B_{\rm s} $ and $ B_{\rm A} $ are the magnetic flux densities at the
stellar surface and at the Alfv\'{e}n radius, respectively.

It is usually assumed that the thermal wind velocity is of the order
of the escape velocity \citep{tout1992,stepien1995} so that
\begin{equation}\label{meq5}
v_{\rm A} \approx \sqrt{\dfrac{2GM}{R_{\rm A}} } .
\end{equation}
Combining equations~\ref{meq5}, \ref{meq4}, \ref{meq3} and~\ref{meq2}
we find equation~\ref{meq1} becomes
\begin{equation}
\frac{dJ_{\rm w}}{dt}=-\left( {(-\dot{M}_{\rm w}})^{(4n-9)} B_{\rm s}^8 (2GM)^{-2} R^{8n}\right)^\frac{1}{4n-5} \Omega.
\end{equation}

The second and considerable torque arises between star and disc by
magnetic interaction. The framework of such an interaction was
constructed by \citet{ghosh1978}.  Some of the stellar magnetic
dipole flux connects to the accretion disc and transports angular
momentum between star and disc. We use the expression given by
\citet{armitage1996} and assume, as did
\citet{stepien2000,stepien2002}, that the radius of magnetosphere is
equal to the co-rotation radius,
\begin{equation}
R_{\rm{cor}} =(\frac{\Omega}{\Omega_k})^{-2/3}R  .
\end{equation}
As they pointed out the disc torque does not depend on the mass of the
disc so
\begin{equation}
\frac{dJ_{\rm disc}}{dt}=-\frac{\mu^2\Omega^2}{3GM},
\end{equation}
where $\mu=B_{\rm s}R^3$ is the magnetic moment of stellar magnetic
field.  Contrary to \citet{stepien2002} who assumes magnetic flux is
constant, we assume that the magnetic field strength $B_{\rm s}$
remains constant because the mass of the accreting star increases at a
substantial rate ($ \dot{M}_{\rm acc} \approx 10^{-5}\,{\rm
  M_\odot\,yr^{-1}} $).  We seek a solution for various fiducial
values of $B_{\rm s}$.

The third torque, the accretion torque discussed in
section~\ref{models}, has the effect of spinning-up the mass-accreting
star.  Now we replace the radius of the star $R$ in
equation~\ref{Tacc} with $R_{\rm{cor}}$ because the stellar magnetic
field disrupts the disc at $R_{\rm{cor}}$ as constant at the value of
co-rotation radius.

The wind mass-loss rate remains an unknown parameter.  Although there
is no a priori rate we can set an upper limit.  All observed Algols
show reversed mass ratio so much of the material lost by the donor
must be accreted by the gainer.  We may write
\begin{equation}
\dot{M}_{\rm{acc}}\approx \beta \dot{M}_{2} 
\end{equation}
and
\begin{equation} \label{eqmw}
\dot{M}_{\rm w} \le (1-\beta) \dot{M}_{2} = \dfrac{(1-\beta)}{\beta} \dot{M}_{\rm{acc}},
\end{equation}
where $0<\beta <1$.  For conservative evolution $\beta = 1$.
\citet{matt2005} claimed that the mass outflow rate in stellar winds
is about 10\,per cent of the accretion rate for the pre-mainsequence
stars, corresponding to $\beta \approx 0.9$.  This is why the
classical T~Tauri stars spin at less than 10\,per cent of their
breakup velocity.  It is the original suggestion by \citet{hartmann1989}
that about one tenth of accreted material lost in a wind can remove
the accreted angular momentum.

Such models have already been applied to a wide variety accreting
stars including the pre-mainsequnce stars
\citep{matt2005,matt2008a,matt2008b} and the Ap and Be stars
\citep{stepien2000,stepien2002}.  Both $\dot{M}_{\rm{acc}}$ and
$\dot{M}_{\rm w}$ are taken as free parameters in most of these
studies.  In the case of binary evolution $\dot{M}_{\rm{acc}}$ depends
on the orbital evolution which can be well modelled.  So, for a given
magnetic field, we need only make an estimate of the fraction of mass
lost in the wind (equation~\ref{eqmw}) to estimate the angular
momentum loss.

\begin{figure*}
\includegraphics[width=17 cm]{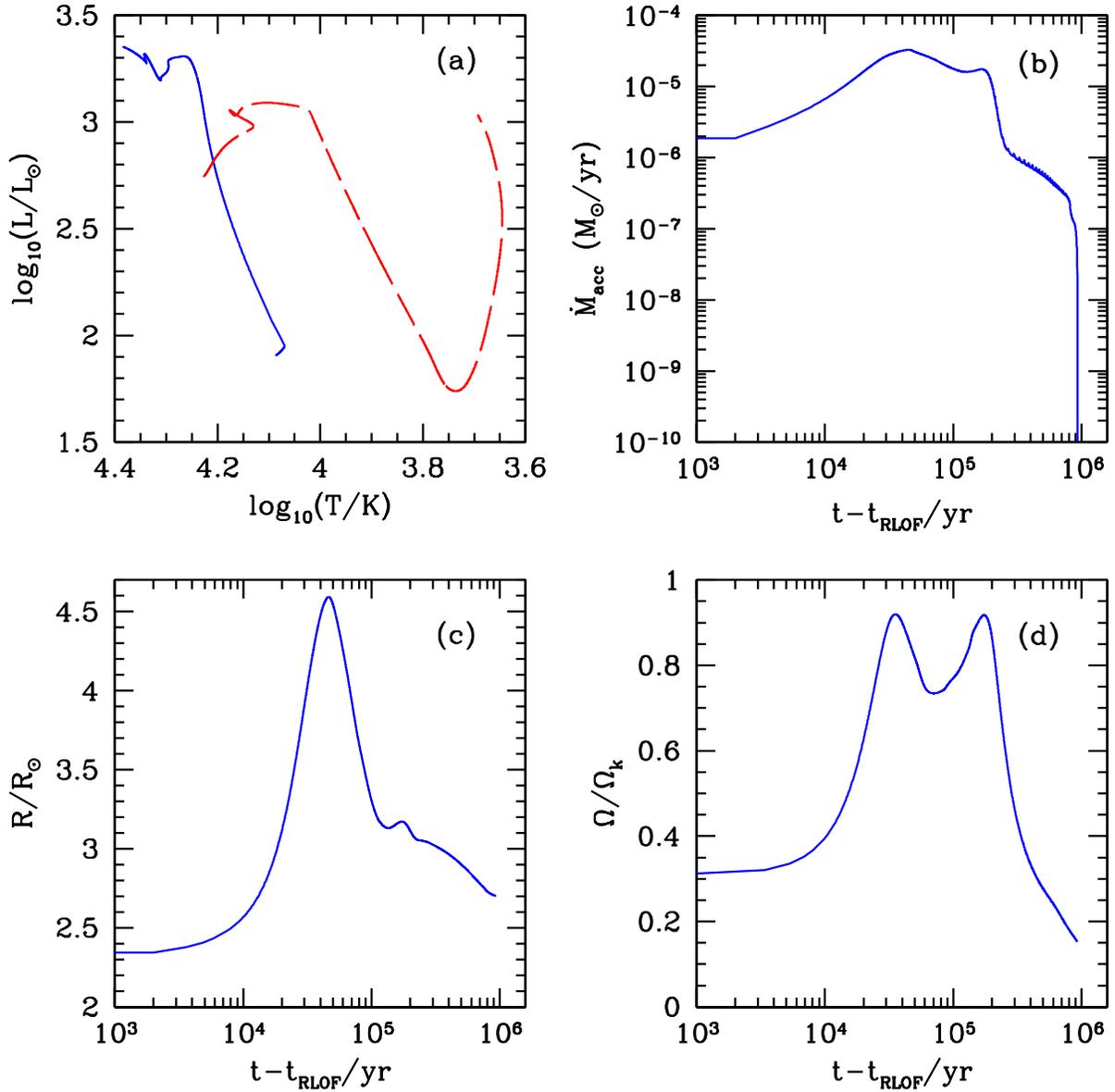}
\caption{Evolution of a binary star with initial mass of $5$ and
  $3\,\rm{M}_{\odot}$ and orbital period of $5\,$d with $\beta = 0.9$
  and $B_{\rm s}=1.5$\,kG.  In (a) the dashed line is the track of the
  initially more massive donor and the solid line that of the gainer
  in an Hertzsprung--Russell diagram.  Panels (b), (c) and (d) show
  the accretion rate, radius and $\Omega/\Omega_k$ of the gainer as a
  function of time since the onset of Roche lobe overflow
  (RLOF).\label{evol1}}
\end{figure*}

Like the tidal torque, which also depends strongly on the separation of
binary, the magnetized wind and disc torques are very sensitive to
changes in the radius $R$ of the mass accreting star so a detailed
evolutionary model is required.  Here we use binary star
models evolved with the Cambridge STARS code which was originally
developed by \citet{egg1971}.  The physics was systematically updated
by \citet{pols1995} and the code has been modified so that it can
evolve both components of a binary system simultaneously together with
the effects of mass and angular momentum loss and transfer
\citep{stancliffe2009}.

We have made various binary evolution models for two different initial
systems, each with a period of $5\,$d and with masses
$5+3\,\rm{M}_{\odot}$ and $3.2+2\,\rm{M}_{\odot}$.  Such systems are believed
to be the progenitors of Algol-type systems.  We evolved both
with $\beta = 0.9$, 0.7, 0.5 and~$0.1$.  The code does not directly
take into account the presence of a disc so we tabulate the output of
the one dimensional structure of each component, the mass-transfer
rate, accretion-rate and orbital parameters at every time step.  From
this we calculate the torque of the magnetic wind, the disc and the
accretion and hence the resulting rotation rate for a given magnetic
field strength.  Because the variation in radius of the gainer is
taken directly from the tables it is an approximation.  It does
not take account of possible rapid rotation.  Typically
the radius of an early-type non-rotating star depends on the mass as
$R \approx M^{0.67}$ at solar metallicity \citep{maeder2009}.  A
rapidly rotating star becomes oblate.  Its polar radius becomes
smaller than that of the equatorial radius so that the equatorial
radius, when rotating solidly at critical velocity is increased by a
factor of 1.5 with respect to a non-rotating star of the same mass
\citep{ekstrom2008}.  However when $\Omega/\Omega_k \approx 0.8$ this
factor is only~1.1 and can be neglected when compared to
the increase in radius as matter accretes.

To calculate the spin evolution of the mass accreting component we assume a
dipolar $(n=3)$ magnetic field and that a magnetized wind is launched as
soon as mass transfer begins.  We first obtain an evolutionary model
of the given binary and each value of $\beta$.  Then for various
field strengths of $B_{\rm s}$ for each $\beta$ we examine carefully how the
angular velocity of the star changes as its mass increases.  If the
star never reaches its critical velocity the computation is ended.
If, on the other hand, the star reaches its critical rotation rate we
interrupt the mass accretion but allow the magnetic wind and disc
torques to lower the angular velocity to less than the critical rate.
This leads to an extra mass loss and a time lag in the evolution of
the system. When the gainer loses enough angular momentum that
$\Omega/\Omega_k < 1$ mass accretion resumes.

In Fig.~\ref{evol1} we show the evolution of the system with initial
masses~$5$ and~$3\,\rm{M}_{\odot}$ and orbital period of~$5\,$d.  The
$5\,\rm M_\odot$ star evolves off the main sequence when it exhausts
hydrogen in its core and moves towards the Hertzsprung gap where it fills its
Roche lobe and rapidly transfers most of its envelope to its
less massive companion.  It becomes established as a red giant when
its mass has fallen to $0.67\,\rm{M}_{\odot}$.  Our models show that
this final mass is almost independent of $\beta$
($0.675\,\rm{M}_{\odot}$ for $\beta=1$ falling to
$0.665\,\rm{M}_{\odot}$ for $\beta=0.5$).  Meanwhile, the gainer moves
up the main sequence as mass transfer proceeds.  In
Fig.~\ref{evol1}a the dashed line is the evolutionary track of the
$5\,\rm M_\odot$ and the solid line that of the $3\,\rm{M}_{\odot}$
star.  The mass accretion rate $\dot{M}_{\rm acc}$ for $\beta = 0.9$
is plotted against time in Fig.~\ref{evol1}b.  It is larger during the
early stages and slows sharply after $10^5\,$yr as the loser reaches
the giant branch.  The gainer reaches hydrostatic equilibrium on a
dynamical time-scale and then thermodynamic equilibrium on the
Kelvin--Helmholtz time-scale.  The variation of the accreting star's
radius over the same time interval is plotted in Fig.~\ref{evol1}c and
its angular velocity in Fig.~\ref{evol1}d when the magnetic field
strength is $B_{\rm s}=1.5\,$kG.  Recall that the variation of angular
velocity is caused by the interaction between the torques of the wind,
the accretion and the disc together with radius changes of the star.
After the phase of rapid mass transfer ($q \approx 0.2$) the slowing
torques dominate.  The location of the donor in the H--R diagram (see
Fig.~\ref{figalgolhr}) shows that it is still at its minimum
luminosity reached following mass transfer.  For this system, this phase
corresponds to an effective temperature of $\log_{10} (T_{\rm
  eff}/{\rm K}) = 3.66$ and luminosity of $\log_{10} (L/{\rm L_\odot})
=2.25$.

\begin{figure*}
\begin{centering}
\includegraphics[width=8 cm]{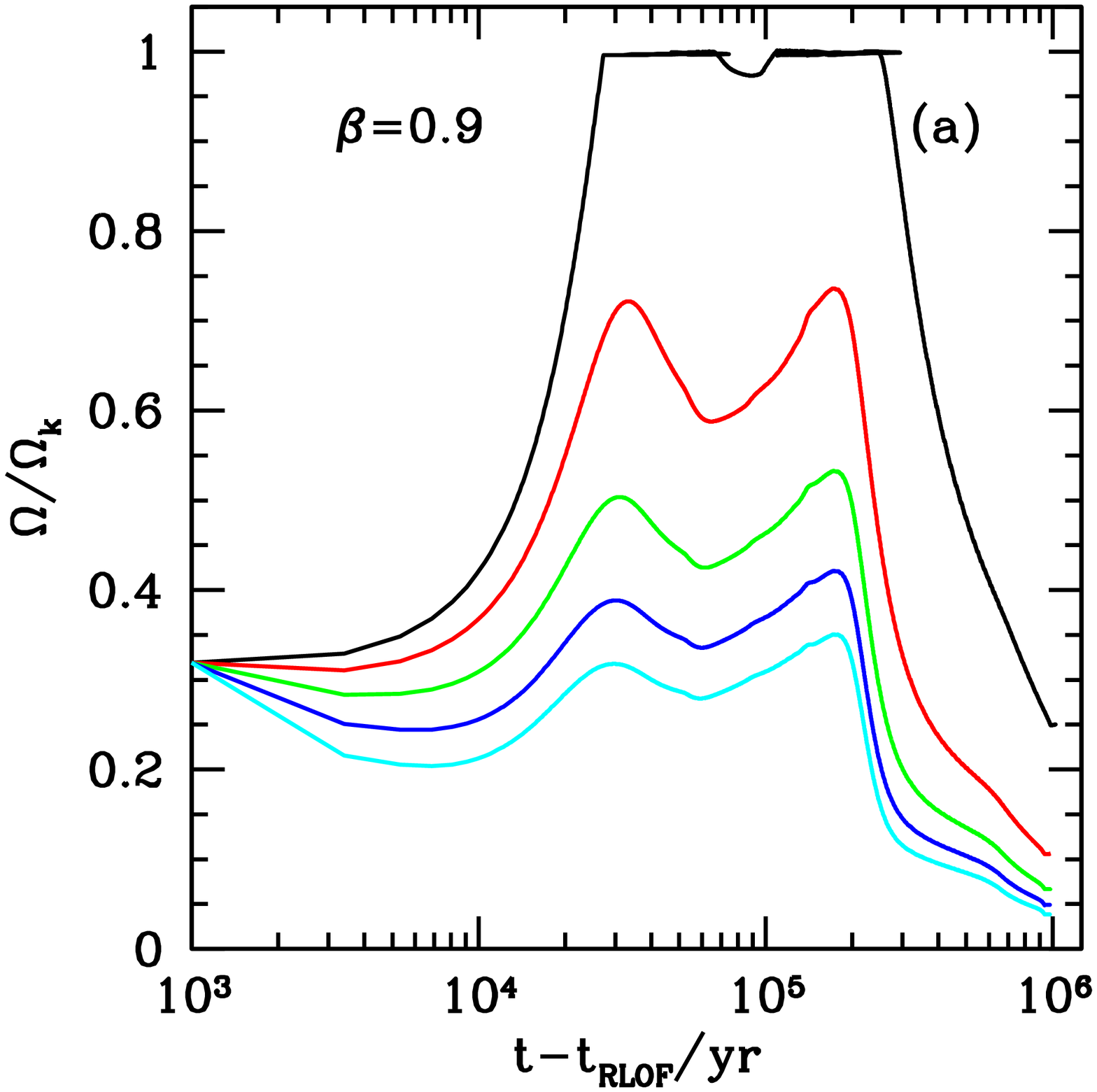}
\includegraphics[width=8 cm]{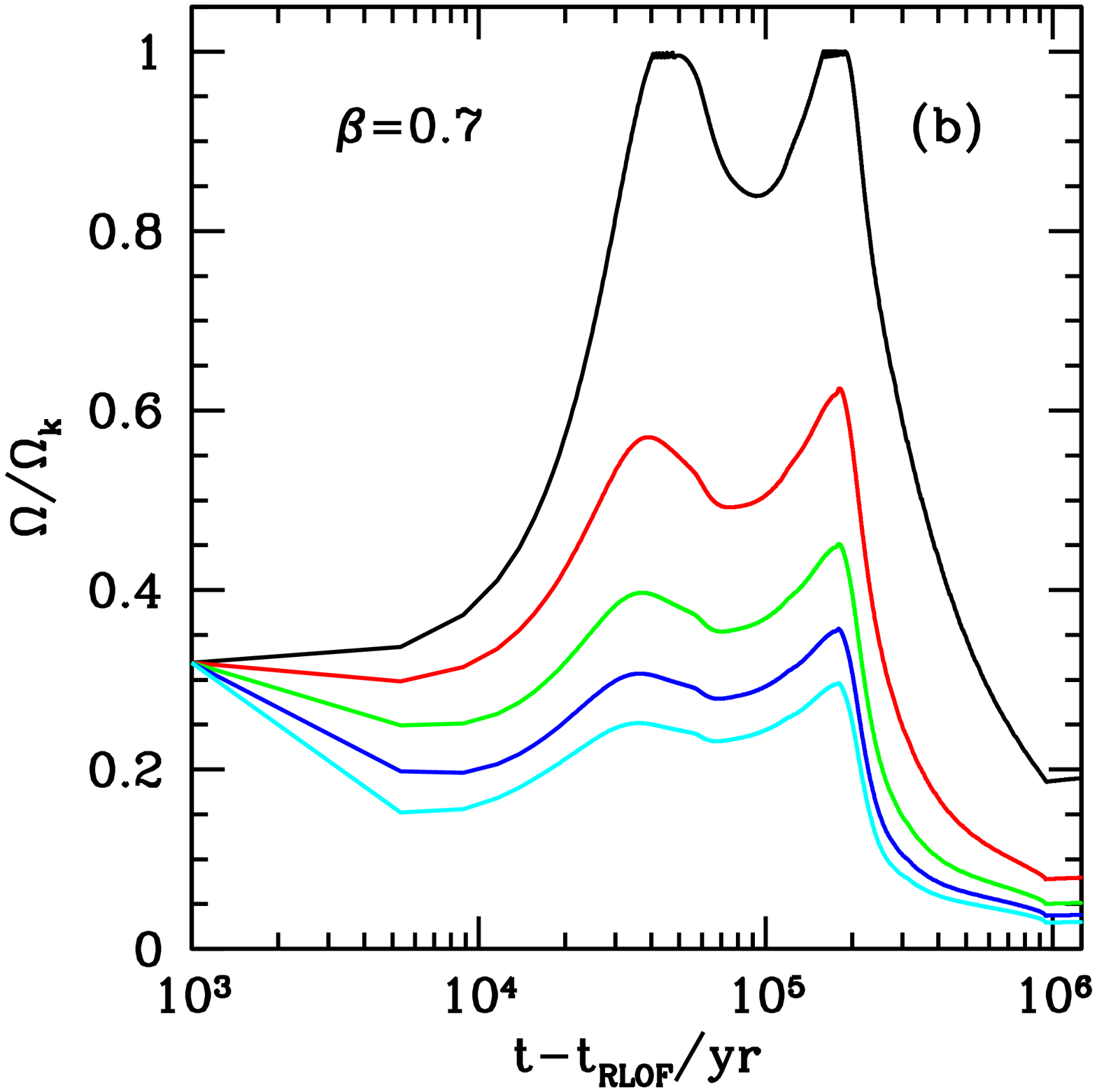}
\includegraphics[width=8 cm]{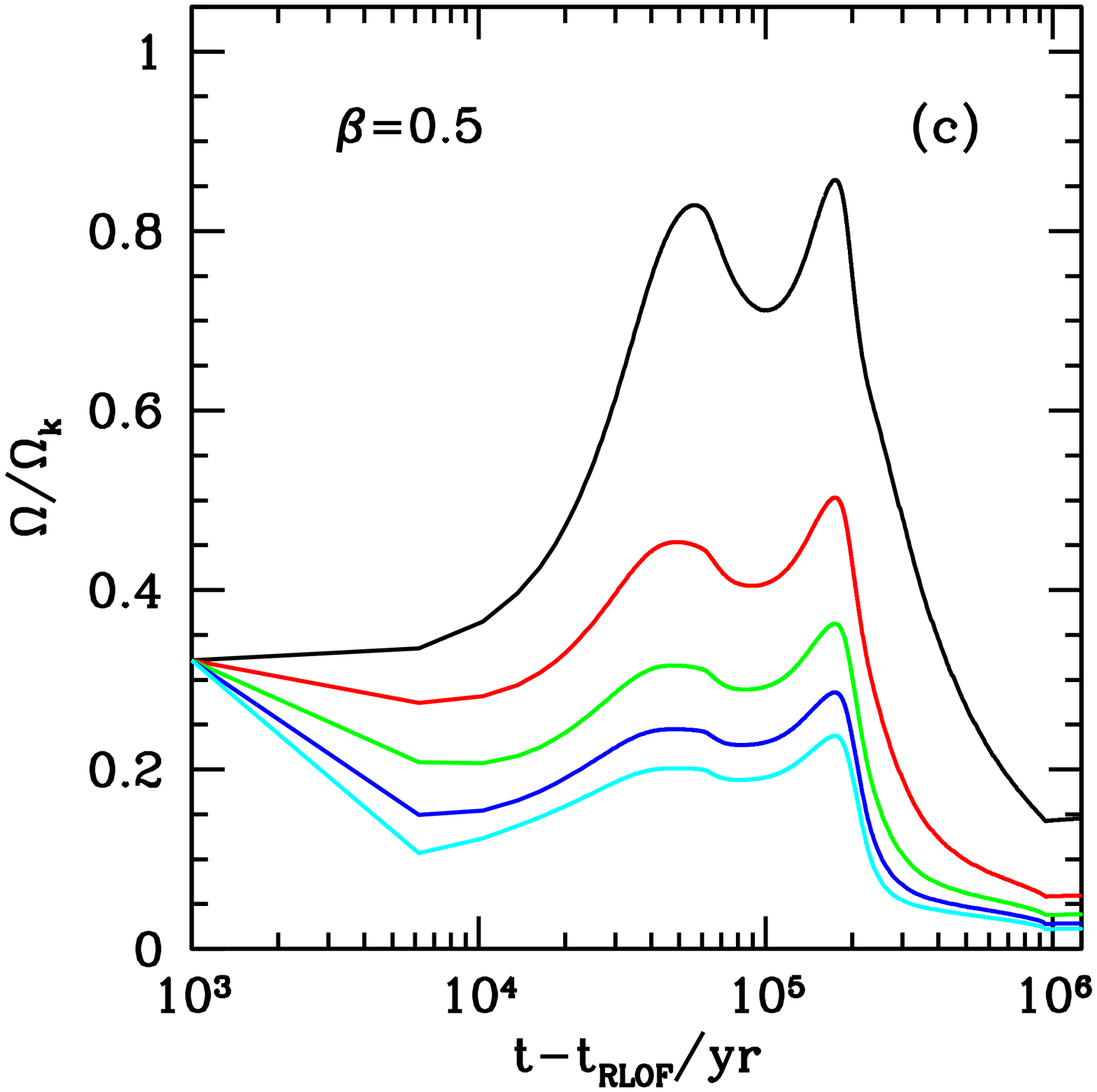}
\includegraphics[width=8 cm]{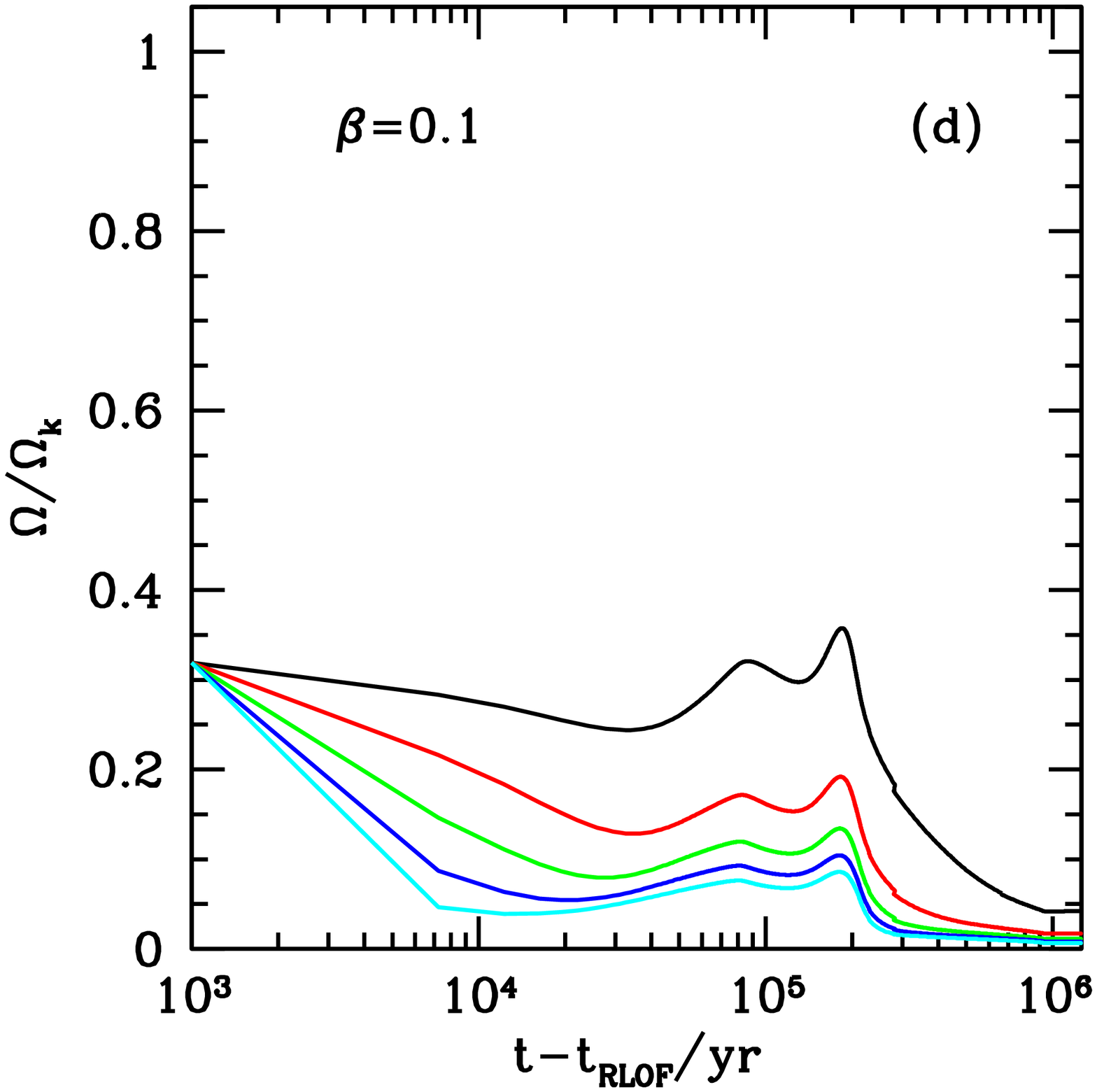}
\par \end{centering}
\caption{Variation of $\Omega/\Omega_k$, the fractional angular
  velocity of a gainer of initial mass $3\rm{M}_{\odot}$, with
  time elapsed since the onset of RLOF for (a)~$\beta = 0.9$, (b)~$\beta = 0.7$, (c)~$\beta = 0.5$
  and (d)~$\beta = 0.1$ with  $B_{\rm s}=1$, 2, 3, 4 and~$5\,$kG, from top to
  bottom in each panel. \label{figB1}}
\end{figure*}

The angular velocities computed for the same initial system but
varying $\beta$ and $B_{\rm s}$ are shown in Fig.~\ref{figB1}.  As
expected a larger $B_{\rm s}$ creates larger torques for the same
$\beta$ and hence smaller angular velocities $\Omega/\Omega_k$ at a
similar point in the evolution.  For the same $B_{\rm s}$ but smaller
$\beta$ there is more mass lost in the wind and again smaller
$\Omega/\Omega_k$ but then the evolution of the system is highly
non-conservative.  The most important result we find here is that when
$B_{\rm s}\gtrsim 1\,$kG the gainer does not reach its critical spin
rate if it loses 10\,per cent of the mass transferred from its
companion.  In such a case a classical Algol as observed today is
produced.  For decreasing $\beta$ the final mass ratio is larger for
the same initial mass.

\begin{figure*}
\includegraphics[width=17 cm]{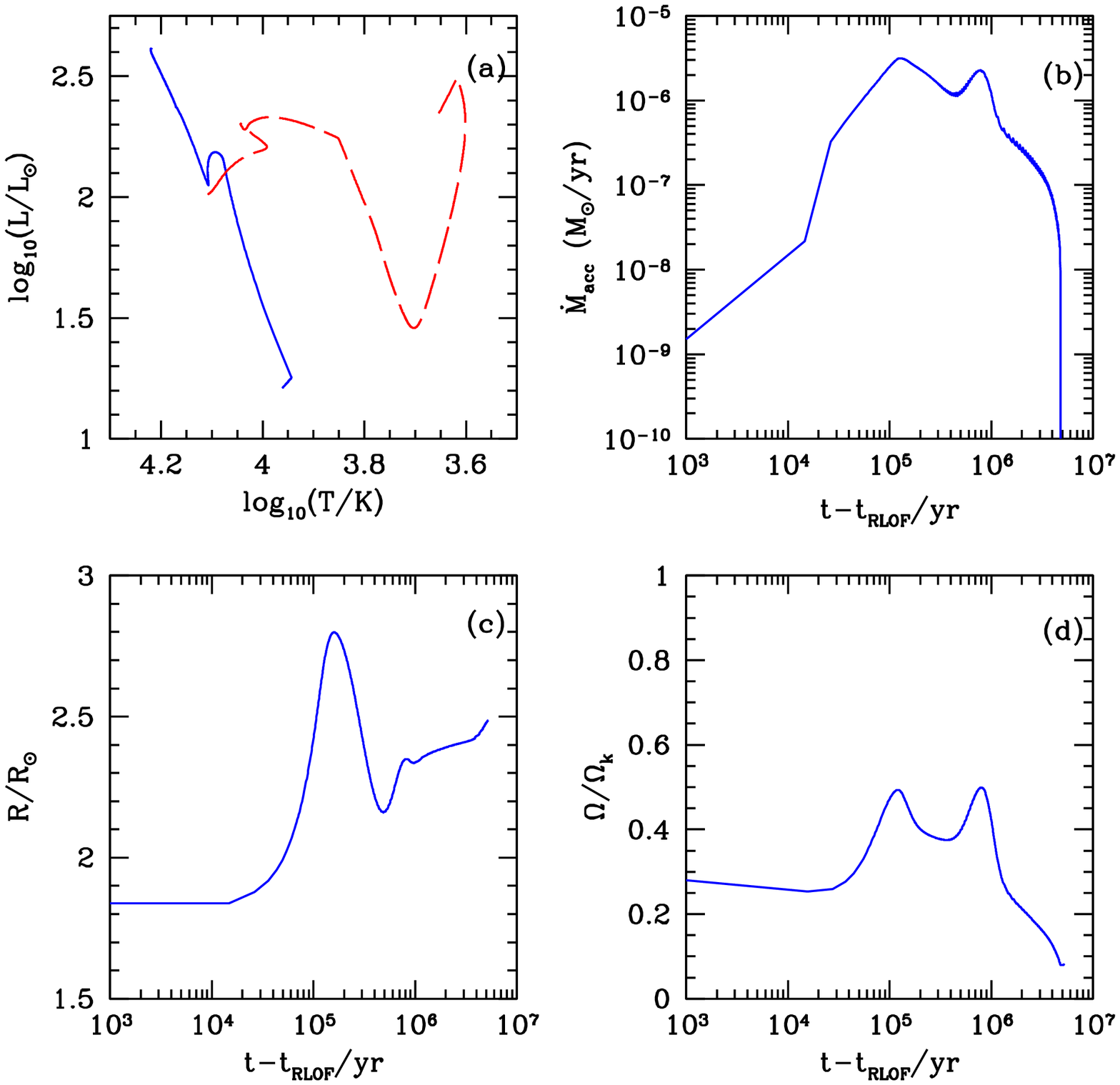}
\caption{Similar to Fig.~\ref{evol1} but evolution of a binary star
  with initial masses of $3.2$ and $2\,\rm{M}_{\odot}$ and orbital
  period of $5\,$d with $\beta = 0.9$ and $B_{\rm s}=1.5$\,kG.  In (a)
  the dashed line is the track of the initially more massive donor and
  the solid line that of the gainer in an Hertzsprung--Russell
  diagram.  Panels (b), (c) and (d) show the
  accretion rate, radius and $\Omega/\Omega_k$ of the gainer as a
  function of time since the onset of RLOF.  \label{evol2}}
\end{figure*}

\begin{figure*}
\begin{centering}
\includegraphics[width=8 cm]{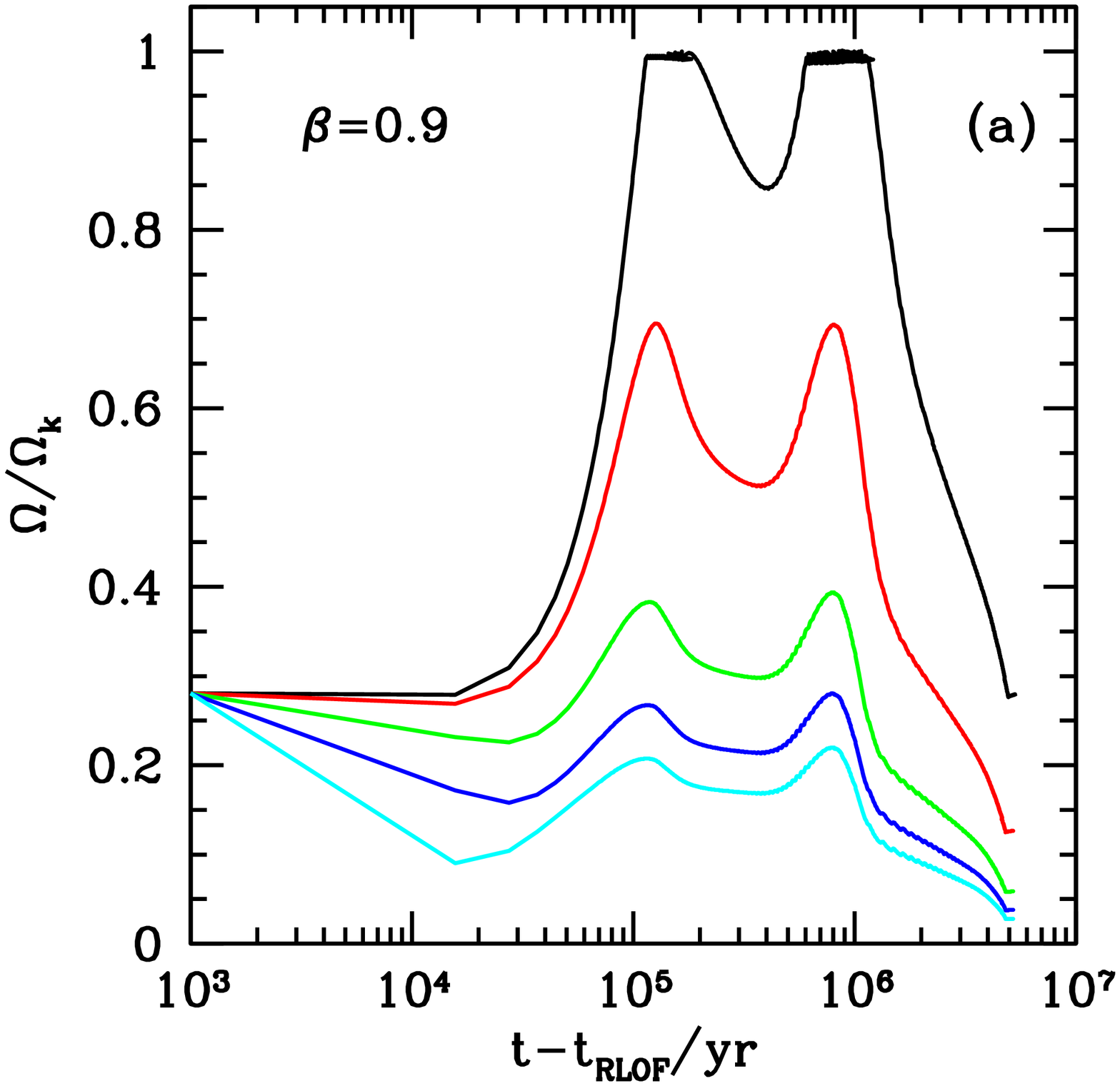}
\includegraphics[width=8 cm]{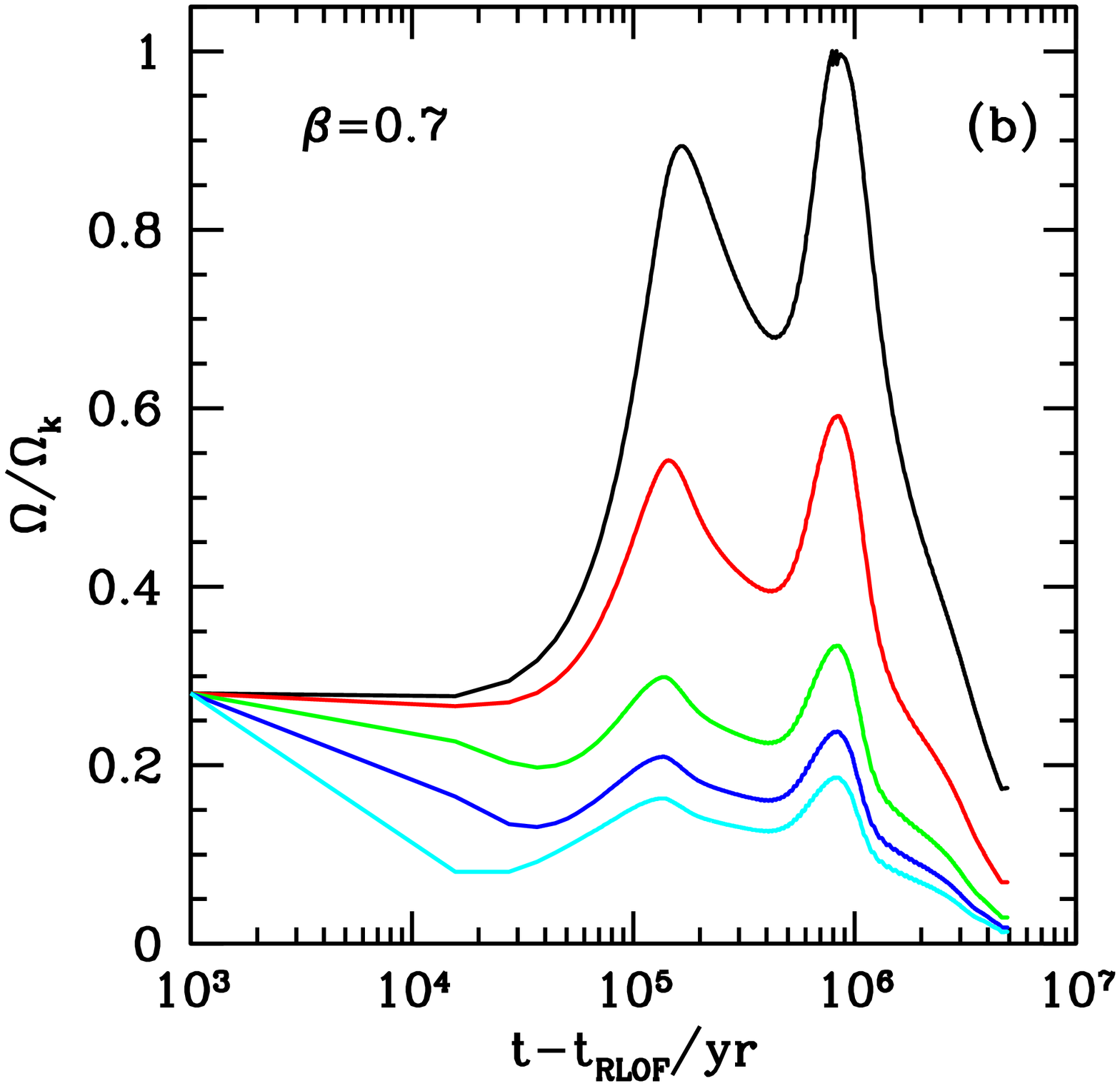}
\includegraphics[width=8 cm]{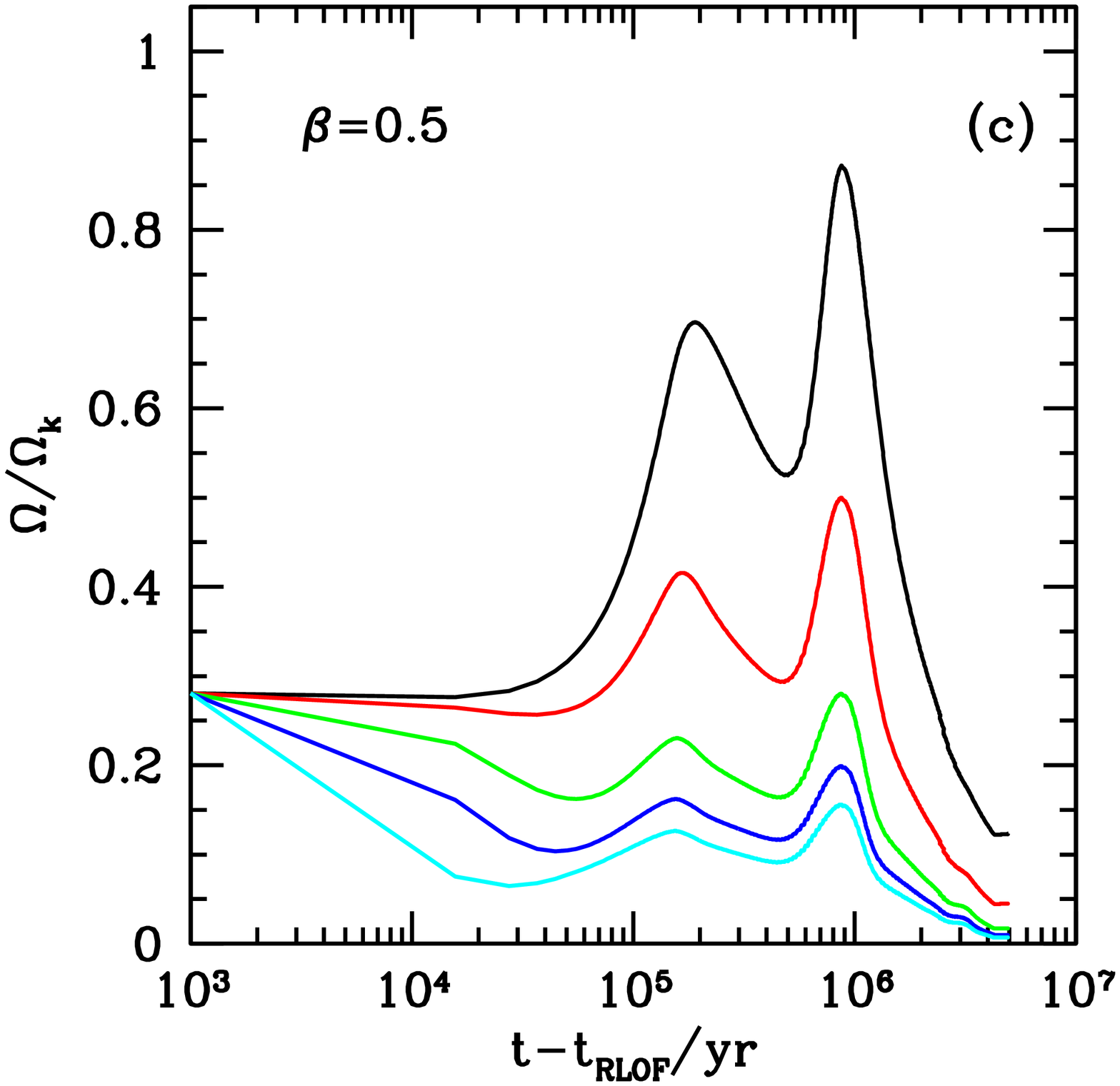}
\includegraphics[width=8 cm]{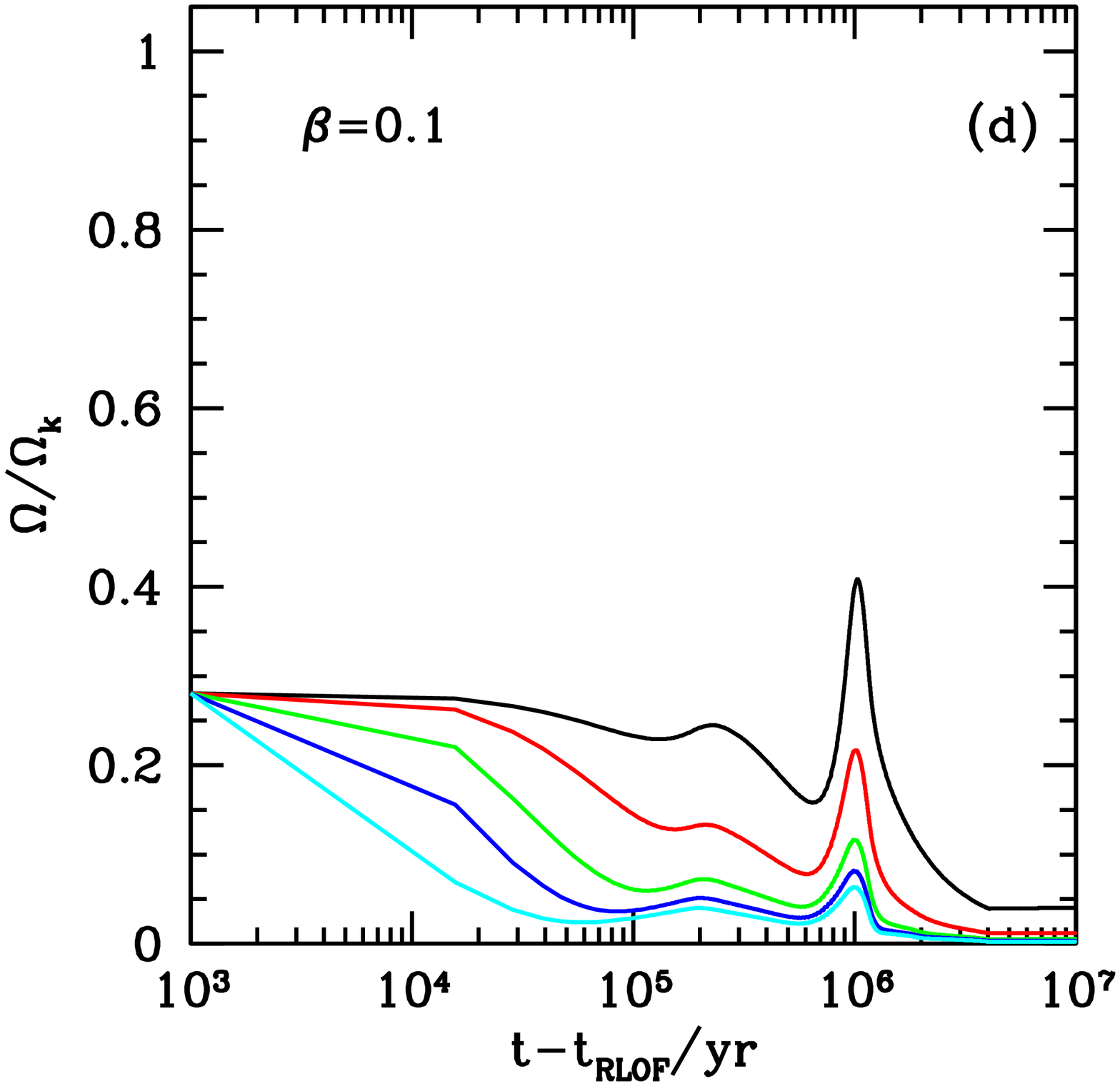}
\par \end{centering}
\caption{ Similar to in Fig.~\ref{figB1} but for initial masses of
  $3.2+2\,\rm{M}_{\odot}$ and period $P=5\,$d, $\beta = 0.9$, 0.7,
  0.5, and 0.1 and $B_{\rm s}=0.5$, 1, 2, 3 and 4\,kG.  \label{figB2}}
\end{figure*}

We also applied our models to a system with initial masses of
$3.2+2\,\rm{M}_{\odot}$ and period $P=5\,$d.  This is typical of the
progenitors considered for Algol ($\beta$~Per) itself
\citep{tout1997}.  The models are shown in Figs.~\ref{evol2} and~\ref{figB2}.

Although there is evidence that magnetic winds brake the spin of
stars, the underlying physics that produces the magnetic fields is not
yet fully understood.  There is a distinction between the magnetic
properties of stars on the upper and lower main sequence.  In the case
of the Sun and solar-like stars the combination of a convective
envelope and rotation is thought to generate the observable magnetic
fields by an $\alpha-\Omega$ dynamo, the strength of which depends
upon the stellar angular velocity.  However, in contrast, only a
minority of early-type stars, such as the Ap and Bp~stars have
observable magnetic fields.  Many attempts have been made to explain
these phenomena \citep[for
  example][]{alecian2007,ekstrom2008,ferrario2009}.  The primaries in
most Algols, especially those with discs, are late~B or early~A type
stars and have radiative envelopes.  Stellar models confirm that the
envelope is convectively stable in massive stars.  Therefore
solar-like concepts of dynamo operation are not directly applicable to
these stars with radiative envelopes.  Furthermore, there is no
observational evidence of the presence of active regions on the
surfaces of A and~B stars.  The origin of any magnetic dynamo which
feeds energy into winds in early type stars is unclear.  Similar
problems also exist for the explanation of the solid-body rotation
profile within the radiative layers of the Sun \citep{chaplin2001}.

While there is a growing body of observational evidence
\citep{donati2006,mm2004} that early type stars have measurable
magnetic fields, no attempt has led to a comprehensive theory which
explains both the origin of and the sustaining mechanism of the
magnetic field.  Two leading possibilities are the fossil field theory
\citep{cowling1945} and the dynamo.  Existing theories of dynamo
generated fields focus on different assumptions but propose that
dynamos operate either in the fully convective cores or in the
differentially rotating radiative envelope.  \citet{mm2004} report
serious difficulties with the core dynamo hypothesis suggested by
\citet{char2001} because the core must create magnetic fields that are
much stronger than equipartition values in order to reach to the
surface.

\citet{spruit1999} argues that a principle ingredient of dynamo
operation in a star is rapid rotation.  This is certainly available in
massive stars.  While the driving mechanism of magnetic field
generation is based on hydrodynamical instabilities in solar type
stars, similar instabilities in the radiative stars may be produced by
the magnetic fields themselves.  Examples are the Parker and Tayler
instabilities \citep{parker1979, tayler1973}.  \citet{spruit2002}
formulated a dynamo mechanism for a differentially rotating star in
which Tayler instability replaces the role of convection in closing
the field amplification loop.  The azimuthal component of the magnetic
field grows until it reaches a critical strength by winding up an
initially very small radial component in a non-convective zone.  At
that point the Tayler instability operates on a very short time-scale,
of the order of Alfv\'{e}n crossing time, to regenerate the radial
field.  The result is a predominantly toroidal field and a closed
dynamo loop.  Recent numerical simulations by \citet{brait2004} and
\citet{brait2006} showed that a stable configuration can be reached
within a non-convective star on an Alfv\'{e}n time-scale from an
arbitrary initial magnetic field.  The Spruit-Tayler mechanism is not
the only one that operates in non-convective material.
\citet{balbus1994} have proposed a magneto-rotational instability
which works well for accretion discs without convection.

\begin{figure}
\includegraphics[width=8.5 cm]{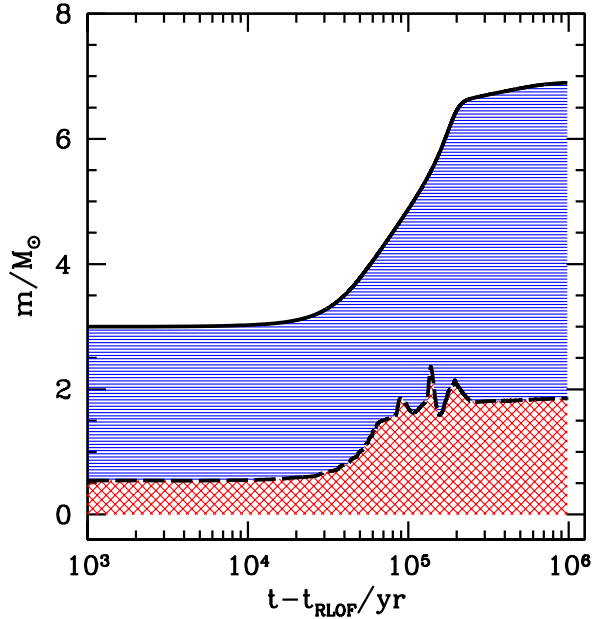}
\caption{Radiative and convective zones with mass $m$ in the gainer in
  our system of initially $5 + 3\,\rm{M}_{\odot}$, orbital period
  $5\,$d and $\beta=0.9$.  Convective (cross shaded) and
  non-convective (horizontal shaded) regions of the gainer during the
  mass transfer are shown with respect to the time elapsed since the
  onset of RLOF.  The envelope of the gainer is always convectively
  stable. \label{figgainer}}
\label{binevol}
\end{figure}

Fig.\,\ref{figgainer} shows the internal structure of the gainer
during the mass transfer for our system with initial masses of $5$ and
$3\,\rm{M}_{\odot}$.  The gainer has a substantial convective core and
an almost constant mass radiative envelope throughout the mass
accretion.

\begin{figure}
\includegraphics[width=8.5 cm]{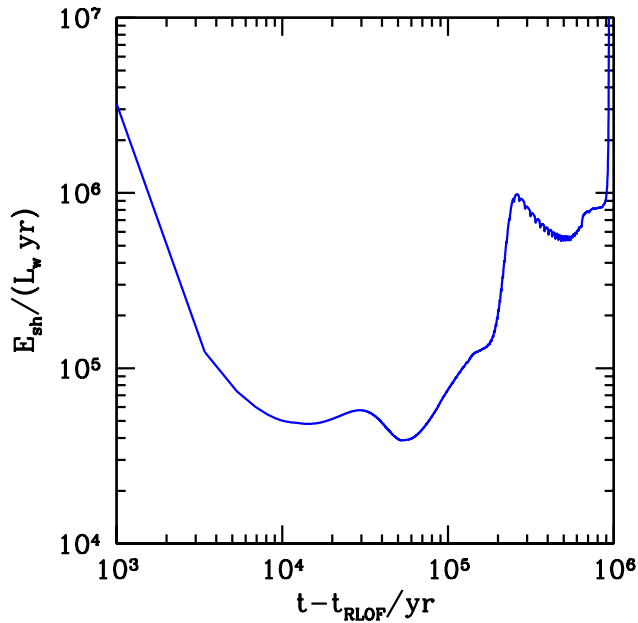} 
\caption{Variation of the dynamo timescale $\tau_{\rm dyn} = E_{\rm
    sh}/L_{\rm w}$ with time elapsed since the onset of RLOF.  The
  shear energy seems to be sufficient to sustain the required wind for
  $10^5 - 10^6 $ years at any stage. \label{figenrat}}
\end{figure}

As \citet{spruit2002} pointed out, differential rotation in single
stars provides a finite amount of energy for generating magnetic fields
so we might expect that the dynamo to cease once the rotation profile
has been smoothed out.  In the case of our semi-detached binary stars
with discs, the energy in differential rotation can be continuously
supplied from the disc material which has high specific angular
momentum relative to the star as we discussed in section~\ref{models}.
Hence, if a Spruit-Tayler type dynamo operates in this case, magnetic
fields are regenerated as long as mass transfer continues.  According
to our assumption that the wind is accelerated to the stellar escape
velocity (equation~\ref{meq5}) and leaves the system at the Alfv\'{e}n
radius, the rate of energy required to be fed into the wind is given
by
\begin{equation}
L_{\rm w} \approx \frac{G M \dot{M}_{\rm w}}{R_{\rm A}}.
\label{lw1}
\end{equation}
This can be provided via the dynamo from
the highly energetic disc material which continuously supplies energy
to the star.  Roughly speaking, $\Delta \Omega \approx R \,
{d\Omega}/{dr}$ is the change of angular velocity between centre and
outer edge of the gainer or across its surface from pole to equator.
The shear energy which can be tapped is \citep{tout1995}
\begin{equation}
E_{\rm sh} \approx \dfrac{1}{2} k^2 \left( \dfrac{GM^2}{R} \right)
\left(\dfrac{\Delta \Omega}{\Omega} \right) \left(
\dfrac{\Omega}{\Omega_{\rm k}} \right) ^2 .
\end{equation}
If $\Delta \Omega \approx \Omega $ we can evaluate $E_{\rm sh}/L_{\rm
  w}$ to estimate the time over which the shear energy could sustain
the wind, even if it were not replenished, to be $10^5 - 10^6\,$yr.
This is similar to the timescale over which mass transfer takes place
so the shear energy is a good candidate for supplying energy to the
dynamo during the mass transfer phase.  The quantity $\tau_{\rm dyn} =
E_{\rm sh}/L_{\rm w}$ during the mass transfer is plotted against the
time since the onset of RLOF in Fig.~\ref{figenrat}.

\citet{tout1992} developed a model based on the idea that the dynamo
process creates not only magnetic field also leads a continual
expulsion of magnetic flux from the star.  This flux, they assume,
provides the mechanical output of energy that drives stellar wind.
They developed schematic dynamo equations and give a rough expression
for this rate as
\begin{equation}\label{lw2} 
L_{\rm w} \approx \dfrac{d}{dt}\left( \dfrac{B_{\phi}^2}{8\pi}
\right)_{\rm loss}\dfrac{4}{3} \pi R^3 \approx
\dfrac{B_{\phi}^2}{4\pi} \tau _{\phi}^{-1}\dfrac{4}{3} \pi R^3,
\end{equation}
where $\tau _{\phi}$ is the decay time-scale of the toroidal field component $B_\phi$.
It is assumed to be equal to $\tau _{\rm g}$, the growth time-scale on
which the
dynamo reaches equilibrium, given by
\citep[Eq~17]{spruit2002}
\begin{equation} 
\tau _{\phi} \approx \tau_{\rm g} \approx \Omega / \omega _{\rm A}^2 \approx q_r^{1/2}.
\label{eq2}
\end{equation}
Combining equations~\ref{lw1} and~\ref{lw2} and using
equation~\ref{eq2} with equation~19 of \citet{spruit2002}, we can
estimate the mass-loss rate in the wind for mean values of $r\approx R$ and $\rho\approx\bar{\rho}$,
\begin{equation}
\dfrac{\dot{M}_{\rm w}}{M} \approx q_r^2\Omega \left(
\dfrac{\Omega}{\Omega _k}\right) ^2 \left( \dfrac{\Omega}{N} \right)
^{1/2}\left( \dfrac{K}{R^2N}\right) ^{1/2} .
\label{mdotw}
\end{equation}
In this formula the quantity $N$ is the Brunt-V\"{a}is\"{a}l\"{a}
frequency, $q_r = d\ln\Omega/d\ln r$ is dimensionless gradient
of rotation and $K$ is the thermal diffusivity.  This equation is derived
under the assumption that almost all the flux lost is transferred to
the wind.  Therefore, equation~\ref{mdotw} is a theoretical upper
limit to the mass lost in the wind.  We find that such a fully efficient
Spruit-Tayler dynamo can support a wind up to
$0.01\,\rm{M_\odot\,yr^{-1}}$.  In this estimate we use mean values of the
parameters $N\approx 10^{-3}\,\rm{s^{-1}}$, $K\approx 10^8-10^9
\rm{cm^2\,s^{-1}}$ and assume the differential rotation gradient
$q_{\rm r}\approx \Delta \Omega / \Omega \approx 1$.
 
\begin{figure}
\includegraphics[width=8.5 cm]{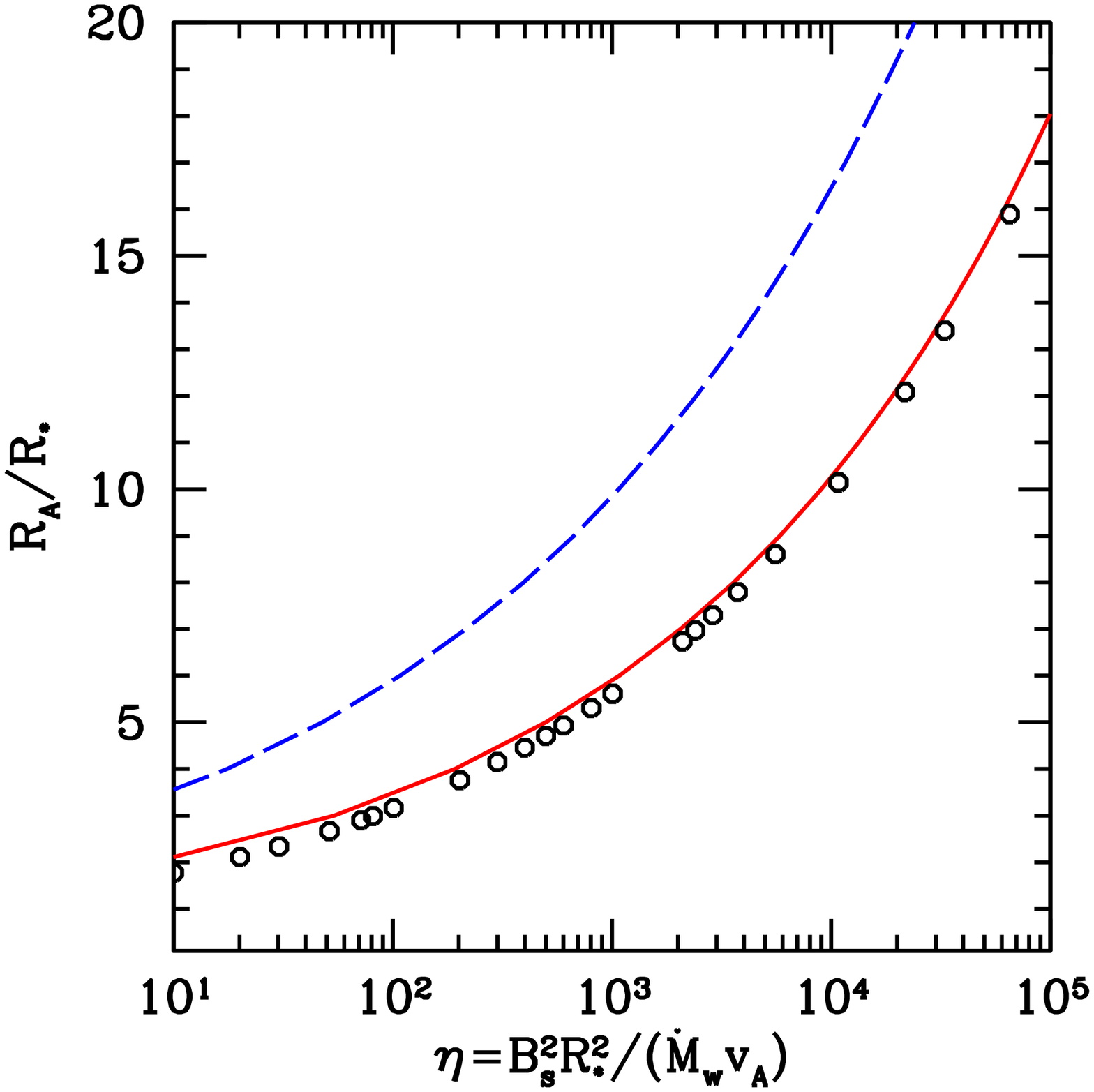} 
\caption{Comparison of our calculated Alfv\'{e}n radii (open circles)
  with those of \citet[][dashed line]{matt2008a} and \citet[][solid
    line]{uddoula2002} for a range of values of the wind confinement
  parameter $\eta$. \label{figmagconf}}
\end{figure}

\citet{matt2005,matt2008a,matt2008b} have also studied accretion
powered stellar winds, both analytically and with two dimensional
magnetic wind solutions.  Although they are interested in T~Tauri type
stars, they claim that their models are independent of dynamo
mechanism.  Based on two-dimensional simulation they found that the
field strength outside the star falls off as if with $n\approx 3.24$
which is very similar to the simple dipolar field decay with $n = 3$.
We compare the values of the Alfv\'{e}n radius we calculate with
those of \citet{matt2008a} and \citet{uddoula2002} in
Fig.~\ref{figmagconf}.  In this comparison we used the dimensionless
wind magnetic confinement parameter, $\eta$ which characterizes the
ratio of magnetic field energy density to the kinetic energy density
of the wind.  We find a good agreement with that of
\citet{uddoula2002}.  However the Alfv\'{e}n radii calculated by us
seem to be a factor of two smaller than those of \citet{matt2008a}.

\section{Conclusions}

It has been known for some time that many long-period Algol systems
have accretion discs.  Accreting material from such a disc should
increase the spin rate of the more massive component up to its
break-up speed as soon as even a small fraction of the mass has been
transferred \citep{demink2007}.  All the classical Algols have a less
massive evolved and a more massive main-sequence component.  Therefore
a substantial amount of mass from the initially more massive star must
be either lost or transferred to its companion.  The angular
velocities of the more massive components in many Algols are measured
with great accuracy.  These measurements show that the gainers rotate
somewhat more slowly than their break-up rates.  Thus there must be a
mechanism which removes spin angular momentum from the rapidly
rotating hot star.  We have demonstrated that tidal interaction
between the components and the orbit is too weak to do this fast
enough.  We show, however, that magnetic braking, driven by a magnetic
dynamo that is maintained by the accretion itself, can remove
sufficient angular momentum.  \citet{mm2004} hypothesized that a
dynamo operating in a sheared radiative region can generate flux tubes
which are able to rise the surface of the star.  A shear dynamo in the
extended radiative envelope of a massive star can therefore serve as
an efficient supplier of magnetic flux to the surface of the star.  We
propose a similar self-consistent model in which a dynamo operates in
shear-unstable material throughout the radiative envelopes of the
gainers in Algols with orbital periods longer than $4-5\,$d. Such
stars can lose angular momentum in magnetized stellar winds, leading
to non-conservative angular momentum evolution.

Under this non-conservative evolution, an Algol system continues to
evolve with relatively little mass loss at a rate $\dot{M}_{\rm
  w}\approx 0.1 \dot{M}_{\rm acc}$ but corotating to a relatively
large Alfv\'en radius $R_{\rm A}$.  When it reaches the classical
Algol phase the less massive component is observed to be more evolved
than the massive one.  Spruit-Tayler instabilities seem most likely to
be responsible for the operation of a magnetic dynamo in massive stars
with radiative envelopes \citep{mm2005}.  However we need to construct
two dimensional models with an accretion disc to test how such a
dynamo really operates in Algols.  Such numerical models could also
supply more realistic Alfv\'{e}n radii based on the field geometry.
Using such detailed models we should be able to find an equilibrium
spin-ratio which can explain the observations in Fig~\ref{figprot}.
We might also take into account the effect of wind induced
hydrodynamic instabilities suggested by \citet{lignieres1996} which
may also effect the differential rotation parameter.

Using a very simple model we computed the spin angular momentum
evolution of gainers with discs in two systems with an initial orbital
period of $5\,$d but different masses.  The results show that even a
small amount of mass, about 10\,per cent of the transferred material,
lost by the gainer with a magnetic field of $1\,$kG is sufficient to
slow down the star to below its break-up velocity.  As the magnetic
field strength increases the rotational velocity of the star decreases
for the same amount of mass loss.

\section*{Acknowledgements}
AD is grateful to the Institute of Astronomy, Cambridge for
hospitality and the Scientific and Technological Research Council of
Turkey (T\"{U}B\.{I}TAK) for PhD fellowships.  CAT thanks Churchill
College for his Fellowship and is very grateful for the hospitality of
IUCAA, Pune and the University of Canterbury where some work on this
manuscript was carried out.  This work has been supported in part by
(T\"{U}B\.{I}TAK) through grant no 109708.  We thank the referee for
his or her helpful comments for improving this manuscript.

\label{lastpage}

\end{document}